\input harvmac.tex
\input epsf 

\newcount\figno
\figno=0
\def\fig#1#2#3{
\par\begingroup\parindent=0pt \leftskip=1cm\rightskip=1cm\parindent=0pt
\baselineskip=11pt
\global\advance\figno by 1
\midinsert
\epsfxsize=#3
\centerline{\epsfbox{#2}}
\vskip 12pt
{\bf Figure \the\figno:} #1\par
\endinsert\endgroup\par
}
\def\figlabel#1{\xdef#1{\the\figno}}
\def\encadremath#1{\vbox{\hrule\hbox{\vrule\kern8pt\vbox{\kern8pt
\hbox{$\displaystyle #1$}\kern8pt}
\kern8pt\vrule}\hrule}}

\def\ntwo{${\cal N}=2$}
\def\none{${\cal N}=1$}
\def\susic{supersymmetric}
\def\susy{supersymmetry}
\def\integer#1{\lfloor #1\rfloor}
\def\RL{{\rm Re}\,}
\def\IM{{\rm Im}\,}

\lref\barb{J. L. F. Barbon, {\it Rotated Branes and N=1 Duality},
Phys.Lett. B402 (1997) 59, hep-th/9703051.}

\lref\brone{N. Evans, C. V. Johnson, A. D. Shapere, {\it Orientifolds,
Branes, and Duality of 4D Gauge Theories}, hep-th/9703210.}

\lref\brtwo{J.H. Brodie, A. Hanany, {\it Type IIA Superstrings, Chiral
Symmetry, and $N=1$ 4D Gauge Theory Duality}, hep-th/9704043.}

\lref\brthree{A. Brandhuber, J. Sonnenschein, S. Theisen,  S. Yankielowicz,
{\it Brane Configurations and 4D Field Theory Dualities}, hep-th/9704044.}

\lref\brfour{O.~Aharony, A.~Hanany,
{\it Branes, Superpotentials and Superconformal Fixed Points},
hep-th/9704170.}

\lref\brfive{R. Tatar, {\it Dualities in 4D Theories with Product 
Gauge Groups from Brane Configurations}, hep-th/9704198.}

\lref\brsix{A. Hanany,  A. Zaffaroni, {\it Chiral Symmetry from Type IIA
Branes},
hep-th/9706047.}

\lref\katz{S. Katz, C. Vafa, {\it Geometric Engineering of N=1 Quantum Field
Theories}, hep-th/9611090.}

\lref\sadov{M. Bershadsky, A. Johansen, T. Pantev, V. Sadov, C. Vafa,
{\it F-theory, Geometric Engineering and N=1 Dualities}, hep-th/9612052.}

\lref\vaoo{H. Ooguri, C. Vafa, {\it Geometry of N=1 Dualities in 
Four Dimensions}, hep-th/9702180.}

\lref\vazw{C. Vafa, B. Zwiebach, {\it N=1 Dualities of SO and USp 
Gauge Theories and T-Duality of String Theory}, hep-th/9701015.}

\lref\an{C. Ahn, K. Oh, {\it Geometry, D-Branes and N=1 Duality in Four
Dimensions I}, hep-th/9704061.}

\lref\antwo{C.Ahn, {\it Geometry,D-Branes and N=1 Duality in Four 
Dimensions II}, hep-th/9705004.}

\lref\anthree{C. Ahn, R. Tatar, {\it Geometry, D-branes and N=1 Duality 
in Four Dimensions with Product Gauge Group}, hep-th/9705106.}

\lref\anfour{C. Ahn, K. Oh, R. Tatar {\it Branes, Geometry and N=1 Duality 
with Product Gauge Groups of SO and Sp}, hep-th/9707027.}

\lref\hori{J. de Boer, K. Hori, H. Ooguri, Y. Oz, Z. Yin {\it Branes 
and Mirror Symmetry in N=2 Supersymmetric Gauge Theories in Three
Dimensions}, hep-th/9702154.}

\lref\bru{I. Brunner, A. Karch, {\it Branes and Six Dimensional Fixed Points}
hep-th/9705022.}

\lref\kol{B. Kol, {\it 5d Field Theories and M Theory}, hep-th/9705031.}

\lref\lop{K. Landsteiner, E. Lopez, D. A. Lowe, {\it N=2 Supersymmetric 
Gauge Theories, Branes and Orientifolds}, hep-th/9705199.}

\lref\br{A. Brandhuber, J. Sonnenschein, S. Theisen, S. Yankielowicz,
{\it M Theory and Seiberg-Witten Curves: Orthogonal and Symplectic
Groups.}  hep-th/9705232.}

\lref\altdue {S. Katz, P. Mayr, C. Vafa, 
{\it Mirror symmetry and Exact Solution of 4D N=2 Gauge Theories I}
hep-th/9706110 and references therein.}

\lref\alt{A. Klemm, W. Lerche, P. Mayr, C. Vafa, N. Warner, {\it Self-Dual
Strings and N=2 Supersymmetric Field Theory}, hep-th/9604034.}

\lref\hw{A. Hanany, E. Witten, {\it Type IIB Superstrings, BPS
Monopoles, and Three Dimensional Gauge Dynamics}, hep-th/9611230.}

\lref\witMa{E. Witten, {\it Branes And The Dynamics Of QCD}, hep-th/9706109.}

\lref\witMb{E. Witten, {\it Solutions Of Four-Dimensional Field Theories Via
M Theory}, hep-th/9703166.}

\lref\EGK{S. Elitzur, A. Giveon, D. Kutasov, {\it Branes and N=1 Duality in
String Theory}, Phys. Lett. B400 (1997) 269, hep-th/9702014.}

\lref\EGKtwo{S. Elitzur, A. Giveon, D. Kutasov, E. Rabinovici, A. Schwimmer,
{\it Brane Dynamics and N=1 Supersymmetric Gauge Theory}, hep-th/9704104.}

\lref\DS{M. R. Douglas, S. H. Shenker, {\it Dynamics of $SU(N)$ 
Supersymmetric Gauge Theory}, Nucl.Phys. B447 (1995) 271,
hep-th/9503163.}

\lref\cvj{C. V. Johnson, {\it From M-theory to F-theory, with Branes},
hep-th/9706155}

\lref\spali{A. Fayyazuddin, M. Spalinski, {\it The Seiberg-Witten 
Differential From M-Theory}, hep-th/9706087}
 
\lref\Berkeley{K. Hori, H. Ooguri, Y. Oz, {\it Strong Coupling Dynamics of
Four-Dimensional N=1 Gauge Theories from M Theory Fivebrane},
hep-th/9706082.}

\lref\swone{N. Seiberg, E. Witten, {\it Monopole Condensation And Confinement
In $N=2$ Supersymmetric Yang-Mills Theory }, Nucl.Phys. B426 (1994) 19,
hep-th/9407087.}

\lref\swtwo{N. Seiberg, E. Witten, {\it Monopoles, Duality and Chiral 
Symmetry Breaking in N=2 Supersymmetric QCD}, Nucl.Phys. B431 (1994)
484, hep-th/9408099.}

\lref\yank{A. Brandhuber, N. Itzhaki, V. Kaplunovsky, J. Sonnenschein,
S. Yankielowicz, {\it Comments on the M Theory Approach to N=1 SQCD and Brane
Dynamics}, hep-th/9706127.}

\lref\moroz{A. Marshakov, M. Martellini, A. Morozov, {\it Insights and 
Puzzles from Branes: 4d SUSY Yang-Mills from 6d Models},
hep-th/9706050.}

\lref\dbranerefs{J. Polchinski, {\it Dirichlet-Branes and 
Ramond-Ramond Charges} Phys.Rev.Lett. 75 (1995) 4724, hep-th/9510017.}

\lref\stromtown{A. Strominger, {\it Open P-Branes} Phys.Lett. B383 (1996) 44,
hep-th/9512059;
P. K. Townsend, {\it D-branes from M-branes} Phys.Lett. B373 (1996) 68, 
hep-th/9512062.}

\lref\sun{A. Klemm, W. Lerche, S. Theisen, S. Yankielowicz,
{\it Simple Singularities and N=2 Supersymmetric Yang-Mills Theory},
Phys.Lett. B344 (1995) 169, hep-th/9411048.}

\lref\suntwo{P. C. Argyres, A. E. Faraggi,
{\it The Vacuum Structure and Spectrum of N=2 Supersymmetric SU(N) Gauge
Theory}, Phys. Rev. Lett. 74 (1995) 3931, hep-th/9411057.}

\lref\sunthree{A. Hanany, Y. Oz,
{\it On the Quantum Moduli Space of Vacua of $N=2$ Supersymmetric $SU(N_c)$
Gauge Theories}, Nucl. Phys. B452 (1995) 283, hep-th/9505075.}

\lref\sunfour{P. C. Argyres, M. R. Plesser, A. Shapere,
{\it The Coulomb Phase of N=2 Supersymmetric QCD},
Phys. Rev. Lett. 75 (1995) 1699, hep-th/9505100.}

\lref\gmv{B. R. Greene, D. R. Morrison, C. Vafa, {\it 
A Geometric Realization of Confinement}, Nucl.Phys. B481 (1996) 513, 
hep-th/9608039.}
 
\lref\dvalshif{G. Dvali, M. Shifman, {\it Domain Walls in Strongly Coupled 
Theories}, Phys.Lett. B396 (1997) 64, hep-th/9612128; A. Kovner,
M. Shifman, A. Smilga, {\it Domain Walls in Supersymmetric Yang-Mills
Theories}, hep-th/9706089; B. Chibisov, M. Shifman, {\it BPS-Saturated
Walls in Supersymmetric Theories}, hep-th/9706141.}  

\lref\memIIa{ M.J. Duff, P.S. Howe, T. Inami and K.S. Stelle, 
{\it Superstrings In D = 10 From Supermembranes In D = 11}, Phys. Lett
191B (1987) 70; P.K.  Townsend. {\it The Eleven-Dimensional
Supermembrane Revisited}, Phys.Lett. B350 (1995) 184, hep-th/9501068;
E. Witten, {\it String Theory Dynamics in Various Dimensions},
Nucl.Phys. B443 (1995) 85, hep-th/9503124.}

\lref\tooft{G. 't Hooft, {\it Topology of the Gauge Condition and New
Confinement Phases in Non-Abelian Gauge Theories}, Nucl.Phys. B190
(1981) 455.}

\lref\nonsusybrane{ A. Brandhuber, J. Sonnenschein, S. Theisen, 
S. Yankielowicz, {\it Brane Configurations and 4-D Field Theory
Dualities}, hep-th/9704044; N. Evans, {\it Softly Broken SQCD: in the
Continuum, on the Lattice, on the Brane}, hep-th/9707197.}  

\lref\lattsusy{I. Montvay, {\it Supersymmetric Gauge Theories on the 
Lattice}, Nucl. Phys. B53, Proc. Suppl. (1997) 853-855,
hep-lat/9607035, and references therein.}

\lref\lattrev{See for example E. Marinari, M.L. Paciello, B. Taglienti,
 {\it The String Tension in Gauge Theories},
Int.J.Mod.Phys. A10 (1995) 4265, hep-lat/9503027.}

\lref\NielOle{H.B. Nielsen, P. Olesen, Nucl.Phys. B61 (1973) 45.}

\lref\Bog{E.B. Bogomolny, Sov.J.Nucl.Phys. 24 (1976) 449.}

\lref\AHvort{H.J. de Vega, F.A. Schaposnik, Phys.Rev. D14 (1976) 1100;
J. Edelstein, C. Nunez, F. Schaposnik, {\it Supersymmetry and
 Bogomolny Equations in the Abelian Higgs Model}, Phys.Lett. B329
 (1994) 39, hep-th/9311055.}

\lref\vortexsubtle{
A.A. Penin, V.A. Rubakov, P.G. Tinyakov, S.V. Troitskii, {\it What
becomes of vortices in theories with flat directions}, Phys.Lett. B389
(1996) 13-17, hep-ph/9609257; O. Aharony, A. Hanany, K. Intriligator,
N.  Seiberg, M.J. Strassler, {\it Aspects of \ntwo\ Supersymmetric
Gauge Theories in Three Dimensions}, hep-th/9703110.}

\Title{\vbox{\rightline{hep-th/9707244}
\rightline{IASSNS--HEP--97/91}}}
{\vbox{\centerline{Confinement and Strings in MQCD}}}

\centerline{Amihay Hanany,  Matthew J.~Strassler  and  Alberto Zaffaroni}
\smallskip{\it
\centerline{School of Natural Sciences}
\centerline{Institute for Advanced Studies}
\centerline{Princeton, NJ 08540, USA}}
\centerline{\tt hanany@ias.edu \tt strasslr@ias.edu \tt zaffaron@ias.edu}

\vskip .2in

\vglue .3cm

\noindent

We study aspects of confinement in the M theory fivebrane version of
QCD (MQCD).  We show heavy quarks are confined in hadrons (which take
the form of membrane-fivebrane bound states) for \none\ and softly
broken \ntwo\ $SU(N)$ MQCD.  We explore and clarify the transition
from the exotic physics of the latter to the standard physics of the
former.  In particular, the many strings and quark-antiquark mesons
found in \ntwo\ field theory by Douglas and Shenker are reproduced.
It is seen that in the \none\ limit all but one such meson disappears
while all of the strings survive.  The strings of softly broken \ntwo,
\none, and even non-supersymmetric $SU(N)$ MQCD have a common ratio
for their tensions as a function of the amount of flux they carry.  We
also comment on the almost BPS properties of the Douglas-Shenker
strings and discuss the brane picture for monopole confinement on
\ntwo\ QCD Higgs branches.

\Date{7/97}

\newsec{Introduction}
\seclab{\intro}
Recently, many interesting results about field theory have been
obtained by realizing gauge field theories on the world-volume of
branes in string theory.  A particularly interesting configuration was
constructed in \hw\ to study ${\cal N}=4$ three dimensional gauge
theories. The construction was generalized to study gauge theories
with the same amount of supersymmetry in various dimensions in
\refs{\witMb,\brfour,\bru,\kol}.  The four dimensional \none\ case was
first studied in \EGK\ and further investigated in
\refs{\hori,\barb,\brone,\EGKtwo,\brtwo,\brthree,\brfour,\brfive,\brsix}.
A different approach, which involves encoding gauge theories in type
II geometry, has been studied in
\refs{\katz,\sadov,\vazw,\vaoo,\an,\antwo,\anthree,\anfour}.

With this construction, we can easily realize a pure \ntwo\ or \none\
four-dimensional gauge theory as the low energy limit of a
configuration of branes in the weakly coupled Type IIA theory. A
powerful method to solve aspects of these theories has been presented in
\witMb. If we go to the strong coupling limit of the Type IIA
string theory, the configuration of branes become smooth enough to
allow a semi-classical analysis in M theory. Using this method, the
Seiberg-Witten curve for a large number of gauge theories was found in
\refs{\witMb,\lop,\br}(see also \refs{\moroz,\spali,\cvj}). A different 
approach based on a fivebrane interpretation of some four-dimensional
theories was earlier used in \refs{\alt,\altdue} to solve a very large
family of models.

The brane theory (which we will call MQCD) is by no means identical to
QCD.  It contains, among other things, extra colored Kaluza-Klein
states from the compact $x^{10}$ direction around which it is wrapped.
The possibility of varying the radius $R$ of this compact direction
makes MQCD a one-parameter generalization of QCD.  For \ntwo\
supersymmetry we may take $R$ small enough that these Kaluza-Klein
states are heavy compared with the QCD scale, and all of the
interesting field theory gauge couplings and BPS states are
independent of $R$.  This is not true for \none\
\witMa\ if we want to study strings and confinement.
Still, MQCD can be a useful method for extracting physics that cannot
be computed or easily visualized in the context of ordinary field
theory.  It was shown in \witMa\ that \none\ MQCD has flux tubes and
undergoes spontaneous breaking of its discrete chiral symmetry. The
tensions of the MQCD strings and BPS-saturated domain walls \dvalshif\
were computed, and a number of interesting results (such as the fact
that the MQCD string can end on a domain wall) were derived. Our point
of view, following \witMa, is to assume that \none\ MQCD is in the
same universality class as QCD and, therefore, has the same
qualitative properties. We will extract some qualitative insights and
one quantitative formula, whose reliability we cannot prove but which
we find suggestive.  This will be extensively discussed in section 9.

In this paper we want to investigate various aspects of confinement in
MQCD.  It will be relatively straightforward, using the results in
\witMa, to introduce heavy quarks into the theory and study the
topological objects corresponding to mesons and baryons. We obtain a
picture which is consistent with the standard lore of confinement in
ordinary QCD.  In addition we consider the possible existence of
multiple stable QCD strings.  In principle, QCD flux tubes can carry
between $1$ and $N-1$ flux units; we will refer to a string with $k$
units of flux as a ``$k$-string''.  A $k$-string could be important in
the dynamics of a meson built from a quark in the $k$-index
antisymmetric tensor representation and a corresponding antiquark.
But it is a dynamical question as to whether the $k$-string is stable
against decay to $k$ $1$-strings. We will show that in \none\ MQCD
(and also non-\susic\ MQCD) the $k$-strings are all stable.

By contrast, in the \ntwo\ gauge theory softly broken to \none, the
physics is quite different. Using the explicit solution for the \ntwo\
theory \refs{\swone,\sun,\suntwo,\sunthree,\sunfour}, Douglas and Shenker
\DS\ found an exotic spectrum in which quarks in the fundamental
representation form $\integer{(N+1)/2}$ distinguishable mesons. The
Weyl group is broken in this theory and different color components of
the quarks are bound by different strings.  However, as is implicit
in \DS, the $N-1$ strings found there are nothing more than a set of stable
$k$-strings.  We show that MQCD reproduces these result, with string
tensions which agree with those found in field theory.  Also, as we
will show, MQCD provides a convenient picture for the transition from
the softly broken \ntwo\ physics to the more conventional \none\
expectations.

Moreover, we find also that all the fivebrane generalizations of QCD
--- the weakly broken \ntwo, the \none\ and even the
non-supersymmetric proposal of \witMa --- exhibit a common universal
ratio between the tensions of the $k$-strings.  This ratio naturally
agrees with the field theoretical prediction for the weakly broken
\ntwo\ theory \DS.  Unfortunately, far from the \ntwo\ limit, the string 
tensions are not protected from renormalization, and so the MQCD
results are highly questionable.  But it is possible that the {\it
ratios} of tensions are weakly renormalized.  The suggestion that the
$k$-strings are stable may well be correct, and it is also possible
that the quantitative MQCD result is fairly accurate for \none\ QCD or
even for non-supersymmetric QCD.  At the moment there is no data on
ratios of string tensions with which to compare; one requires a
lattice computation using a group larger than $SU(3)$.

The paper is organized as follows.  After a rudimentary overview of
confinement and heavy quarks in section 2, we explain in section 3 how
heavy gauge bosons and quarks appear as membranes in \ntwo\ MQCD.  In
section 4 we turn to \none\ MQCD and show that there are no membrane
states corresponding to isolated heavy quarks. However, heavy quarks
can connect to the MQCD strings identified by Witten \witMa, and M
theory membranes corresponding to mesons and baryons can exist in the
theory.  The meson membranes bear some resemblance to the brane
picture for confinement of monopoles in the Abelian Higgs model found
in \gmv.  We discuss the $k$-strings of \none\ MQCD in section 5.

In section 6 we identify the flux tubes of Douglas and Shenker in
MQCD, and show that their quantum numbers and tensions are given
correctly.  As the \ntwo\ breaking parameter is taken to be large, the
brane construction gives a nice physical picture for the transition
from these flux tubes to those of \none\ MQCD. We discuss the process
by which the order-$N$ different mesons decay to a single one during
this transition. We comment in section 7 on the ``almost BPS''
properties of the flux tubes of \DS, and in section 8 present a brane
picture of monopole confinement along Higgs branches of non-abelian
\ntwo\ QCD.  Finally, in section 9, we discuss our observation that the
ratios of the MQCD $k$-string tensions are independent of the amount
of supersymmetry.  We analyze critically this and the other results
obtained in order to clarify to what extent MQCD predictions can be
trusted in ordinary QCD, and note the possibility of numerical tests.
Section 10 contains a brief conclusion, and the appendices present
some conventions and a computation in non-supersymmetric MQCD.

\newsec{Introduction of Heavy Quarks in QCD}
\seclab{\heavy}

Confinement is usually studied by computing the potential between
static sources (or equivalently the expectation value of a Wilson
loop.)  In M theory the easiest technique is to add dynamical but very
heavy quarks to the theory. 

We remind the reader that it is naively reasonable to talk about a QCD 
string in the context of heavy quarks.  Imagine adding two heavy 
quarks, $U$ and $D$, with masses $m_U, m_D\gg \Lambda_{QCD}$, to a
pure $SU(N)$ gauge theory.  Consider the bound states of a $U$ quark
and $\bar D$ antiquark.  These states carry conserved flavor quantum
numbers and so the ground state is stable.  The lowest lying bound
states are smaller than $\Lambda_{QCD}^{-1}$; they only sample a
region where QCD is weakly coupled, and their spectrum is that of the
Hydrogen atom.  Highly excited states of size $L\gg\Lambda_{QCD}^{-1}$
are subject to the linear QCD potential, whose slope is given by the
QCD string tension $T\sim\Lambda_{QCD}^2$, and have mass of order
$m_U+m_D+ T L$.  If the total energy of the string $TL$ is much
greater than twice the mass of the lighter quark, then the string can
break via quark pair production.  However, this process is slow, as
the energy of the string is spread out over a distance $L$ much
greater than the Compton wavelength $m^{-1}$ of the quarks.  Of
course, it is also possible for any excited state to decay via the
emission of glueballs. These are closed string loops of mass near
$\Lambda_{QCD}$. Such a process is of order $1/N^2$ in the large $N$
limit, and can therefore be controlled.

Thus, as long as we consider states of size $L$ in the
intermediate regime between weak coupling and heavy quark pair production,
\eqn\curve{1 \ll \Lambda_{QCD} L \ll {\min(m_U,m_D)\over\Lambda_{QCD}}\ ,}
the meson can be modeled as a quark and antiquark joined by a QCD
string.

\newsec{Identification of Gauge Bosons and Quarks in \ntwo\ Supersymmetry}
\seclab{\twotheory}

In order to set the groundwork for our study of \none\ \susic\
theories, in which quark states do not exist as independent entities,
we first discuss the unconfined quarks and gauge bosons of
\ntwo\ \susic\ theories.  We omit most technical details (which are
discussed extensively in the original papers \refs{\hw, \witMb}) and
instead provide a light review useful (we hope) to the non-expert.

\fig{\ntwo\ $SU(2)$ gauge theory realized by stretching 
two D4 branes between two NS branes.  The configuration is independent
of spacetime; the coordinates $x^4,x^5,x^6$ are shown. The D4 branes
are located at $v=x_4+ix_5 = \pm \phi$; the string marked $W$ is a
charged vector multiplet of mass $2\phi$. The semi-infinite D4 brane
located at $v=m$ introduces a hypermultiplet $Q$ of bare mass
$m$.}{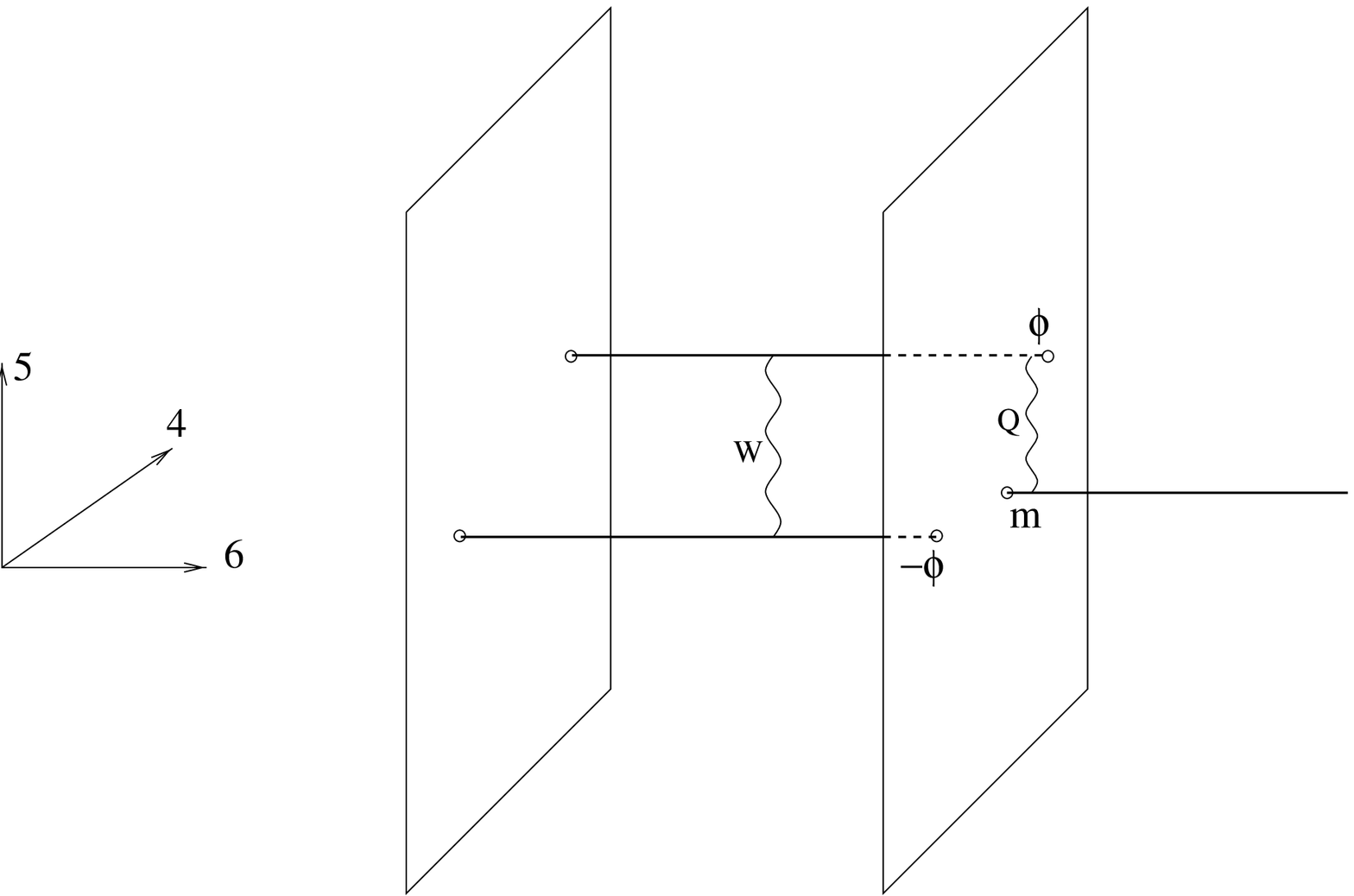}{8 truecm}
\figlabel\sutwo

A simple model studied in Ref.~\swtwo\ is $SU(2)$ gauge theory with a
single hypermultiplet in the doublet representation. 
In Type IIA string theory, the representation of the classical theory
is given in figure \sutwo.  The fields on the world-volume of two
Dirichelet fourbranes (D4 branes) make up a $U(2)$ gauge theory on
five-dimensional Minkowski space $M^5$ \dbranerefs.  We will take the
coordinates of this theory to be $x^0,x^1,x^2,x^3,x^6$.  When the two
D4 branes are stretched between two Neveu-Schwarz fivebranes (NS
branes), which fill coordinates $x^0,x^1,x^2,x^3,x^4,x^5$, this theory
is constrained to exist on $M^4\times I$, where $I$ is an interval of
length $\Delta x^6$.  The NS branes both cut off the volume of the
gauge theory in the $x^6$ direction, making it four-dimensional in the
infrared, and reduce the supersymmetry to the equivalent of \ntwo\ in
four dimensions.  The gauge coupling of the effective four-dimensional
theory on the D4 branes is $1/g^2 = \Delta x^6/g_s \ell_s$, where
$g_s$ and $\ell_s$ are the Type IIA string coupling and length. The D4
branes are free to move in the two dimensions of the NS branes which
are perpendicular to the D4 branes.  These dimensions, $x^4$ and
$x^5$, can be combined into the holomorphic coordinate $v=x^4+ix^5$.
The distance $\delta v$ between the two D4 branes is proportional to
the expectation value $\phi$ of the scalar in the \ntwo\ vector
multiplet of the $SU(2)$ gauge theory; recall this scalar has a single
complex eigenvalue. As explained in \witMb, the requirement of having
finite energy configuration on the NS branes imposes that the sum of
the positions in $v$ of all the D4 branes is zero. As a consequence,
the $U(1)$ subgroup of $U(2)$ is non-dynamical.

The four gauge bosons (and their scalar partners) of the $U(2)$ theory
consist of Type IIA strings which stretch between two (not necessarily
different) D4 branes.  In particular, the $W$ bosons of the
$SU(2)$ subgroup, whose masses are classically proportional to $2\phi$,
are the IIA strings of length $\delta v$ which stretch from the first
D4 brane to the second.  $W^+$ bosons are strings of one
orientation, $W^-$ bosons have the opposite orientation.

A hypermultiplet in the doublet of $SU(2)$ may be added to the theory
by attaching a semi-infinite D4 brane to the right of the righthand NS
brane.  (We will refer to finite D4 branes as color branes, since they
carry color quantum numbers, and to the semi-infinite D4 branes as
flavor branes for the analogous reason.)  The position of the flavor
brane in the $v$ plane is the bare mass for the hypermultiplet.  In
the classical theory, in which $SU(2)$ is broken to $U(1)$ by non-zero
$\phi$, the quark has two color states, of charges $\pm\half$ and
masses $m\pm\phi$.  Since the separations between the flavor brane and
the two color branes are precisely $m\pm\phi$, it is
natural that a quark is a IIA string lying in the NS brane and
stretching from the flavor brane to one of the color branes.  The
string of opposite orientation is the antiquark.

If the Type IIA string coupling is large, the physics of the branes will be
given in terms of the semiclassical limit of the eleven-dimensional
description known as M theory.  This theory exists on $M^{10}\times
S^1$, where the radius $R$ of the circle in the tenth spatial
direction $x^{10}$ grows with the IIA string coupling.  M theory has
two-dimensional membranes and the corresponding electromagnetic dual
objects, fivebranes.  Just as IIA strings can end on D-branes
\dbranerefs, M theory membranes can end on M theory fivebranes
\stromtown, with the intersection being a closed curve inside the
fivebrane.  The IIA string in M theory language is a membrane wrapped
{\it once} around the compact $x^{10}$ direction.  The NS brane is an
M theory fivebrane, while the D4 branes of the IIA string theory are M
theory fivebranes wrapped {\it once} around the compact $x^{10}$
direction.

Thus, the NS and D4 branes of the Type IIA construction outlined above are
made from the same type of object, and it is therefore natural that in
M theory the singular intersections between them would be smoothed
out.  As shown in \witMb\ the construction of figure \sutwo, which is
shown embedded in the space $R^3$ made from $v$ and $x^6$, becomes a
continuous six-dimensional surface filling the {\it eight}-dimensional
space consisting of space-time $M^4$ and the coordinates $v=x^4+ix^5$
and $t = e^{(x^6+ix^{10})/R}$.  Since the construction is
translationally invariant in space-time, the six-dimensional surface
factors into $M^4\times\Sigma$, where $\Sigma$ is a two-dimensional
Riemann surface embedded in the flat $v,t$ space and specified by a
single complex equation in $v$ and $t$.  This Riemann surface is
equivalent to the Seiberg-Witten torus
\refs{\swone,\swtwo} which specifies the gauge coupling of the low-energy
effective $U(1)$ gauge theory.  The embedding of the surface
determines the Seiberg-Witten one-form from which the masses of BPS
states may be determined.

\fig{The curve $\Sigma$ for $SU(2)$ with one flavor; compare with
figure \sutwo.  The curve is wrapped around the compact $x^{10}$
direction, which is not shown.  The intersection of the curve with
$x^{10}=0$ is indicated by the two curved dark lines; notice each
tube corresponding to a D4 brane contains one such line, showing it
wraps once around $x^{10}$.}{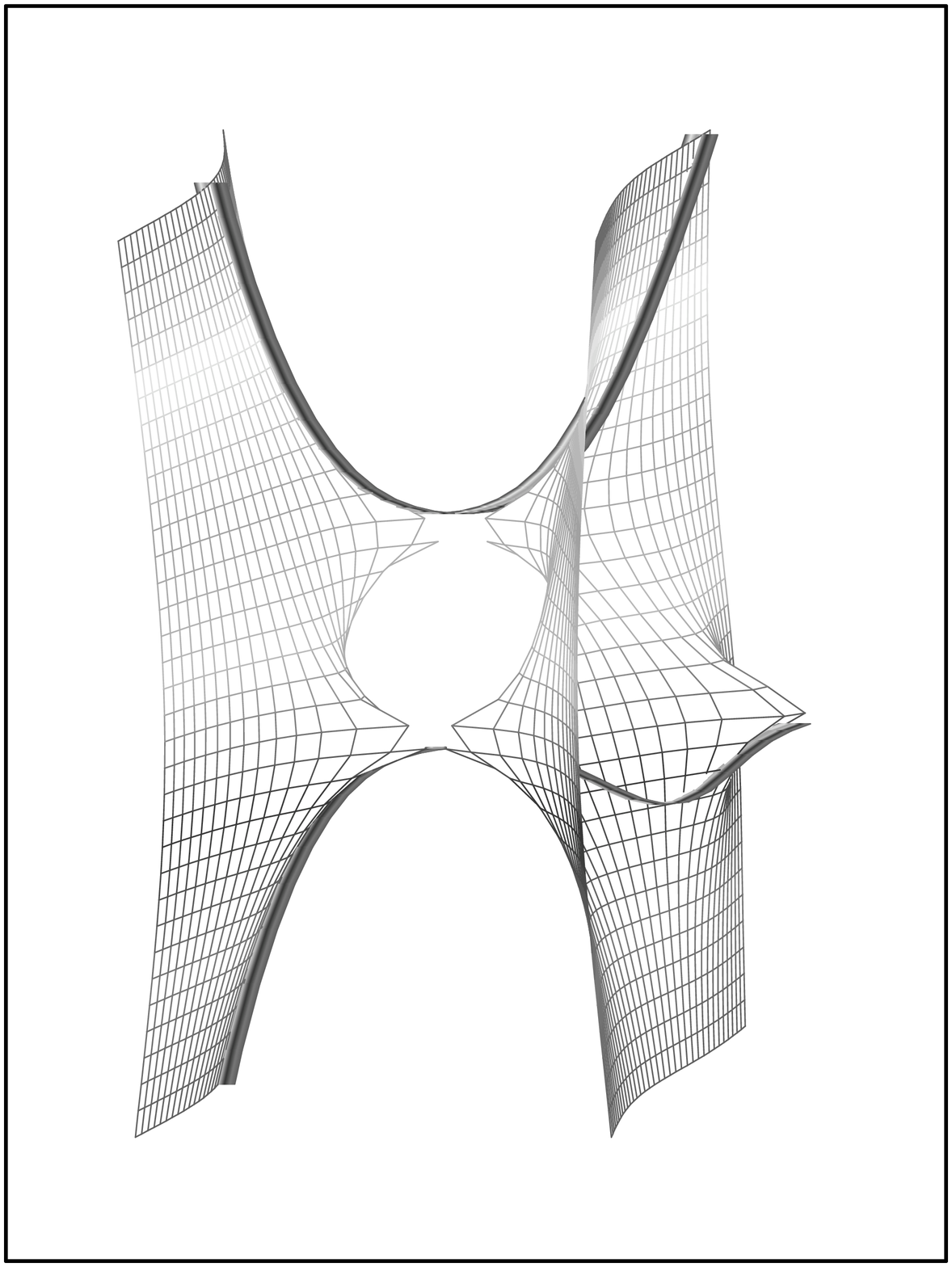}{8 truecm}
\figlabel\sutwobrane
\fig{The curve $\Sigma$ of figure \sutwobrane, projected into the
$v=x^4+ix^5$ plane.  The branch with $|t|>1$ is shown.  The surface
intersects the plane $|t|=1$ along the thick lines, corresponding to
the two colored D4 branes. The contours $\gamma_{\phi}$,
$\gamma_{-\phi}$, $\gamma_m$ are indicated.  The dashed lines
respresent the intersections of the surface with
$x^{10}=0$. }{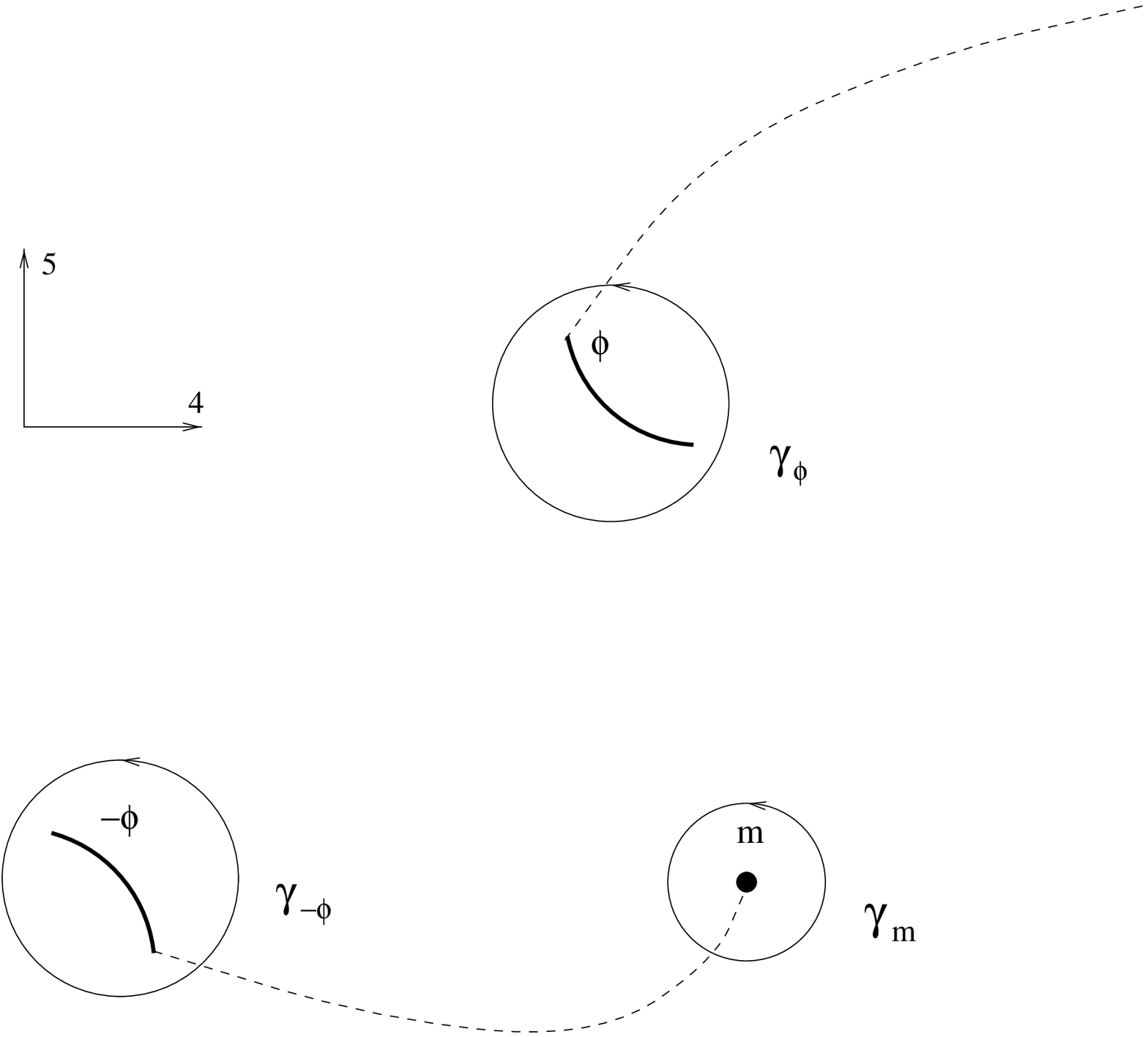}{10 truecm}
\figlabel\due

For the theory in figure \sutwo, the Riemann surface is given by
\refs{\swtwo}
\eqn\curvetwo{
(1-{v\over m})t + \Lambda_2^{-2}(v^2-\phi^2) + 1/t = 0 } 
Note that the gauge coupling has disappeared
and been replaced by $\Lambda_2$.  We show two renderings of the
surface $\Sigma$ \witMb\ in figures
\sutwobrane\ and \due. The first rendering shows the embedding of the 
surface in the $v,|t|$ space.  Although we cannot draw four
dimensions, we note that the surface $\Sigma$ wraps around the compact
direction $\arg t = x^{10}/R$.  We indicate with two dark lines on
$\Sigma$ the points at which $\Sigma$ intersects $x^{10}=0$; a curve
which travels on $\Sigma$ from one dark line to the next wraps once
around $x^{10}$.  Note that the picture (drawn for $\phi\gg\Lambda_2$)
roughly resembles figure \sutwo\ and that, as required, each D4 brane
has become an M theory fivebrane wrapping once around $x^{10}$.  In
figure \due, $\Sigma$ is considered as a double-sheeted cover of the
$v$ plane, with singularities on both sheets near $v=\pm \phi$ and
with a singularity on the top sheet at $v=m$.  We show the top sheet
in the figure.  Here dashed lines indicate intersections of the
surface with $x^{10}=0$.

We now identify the $W$ bosons and quarks in this M theory picture.
The Type IIA strings of figure \sutwo\ which stretch between D4 branes must
now become membranes which wrap once around $x^{10}$ \memIIa\ and
which attach to the fivebrane along closed curves \stromtown.  It is
clear from figure \due\ that the curves
$\gamma_m,\gamma_\phi,\gamma_{-\phi}$ drawn around the three singular
points in $v$ are suitable for the ends of such membranes.  A $W$
boson thus consists of a two-dimensional curve, with cylindrical
topology, lying in $v,t$ but {\it not} in $\Sigma$, which has one
boundary on the contour $\gamma_\phi$ and the other on
$\gamma_{-\phi}$.  Again we emphasize that the boundaries of the
membrane {\it do} lie in $\Sigma$ though the bulk of the membrane does
not.  In fact the membrane will be the minimal area surface in $v,t$
with these boundaries, and will appear roughly as a cylinder of radius
$2\pi R$ and length $2\phi$.  The mass of the four-dimensional
particle is proportional to the area of the membrane, and will
therefore be proportional to $2\phi$.  Similar statements apply to the
two quark states which connect $\gamma_m$ with one of the other two
curves; the masses will be of order $m\pm \phi$.\foot{The details of
dimension counting in this system are given in appendix A.}

As a technical matter, we note that if the theory contained two heavy
quarks, there would be singularities at $v=m_1$ and $v=m_2$, and in
addition to the $W$ boson and quark states, one could consider an open
membrane whose two boundaries wrap around these two singularities.
This would correspond to a gauge boson of the flavor group.  Since the
flavor branes are semi-infinite, the flavor theory is actually five
dimensional and the flavor gauge bosons do not couple dynamically to
the four-dimensional theory.  They instead couple as background gauge
fields to the corresponding flavor currents.

A monopole in Type IIA string theory is a rectangular D2
brane with two boundaries on D4 branes and two on NS branes \hw.  In
figure \sutwo\ the monopole fills the ``hole'' between the branes, like a
soap-bubble.  In M theory the monopole is a membrane stretched across
an opening in the Riemann surface \witMb, such as the large hole in figure
\sutwobrane.  Its mass is proportional to the minimal area of the
hole. When $\phi$ is tuned to a special value, the area of the hole
shrinks to zero, corresponding to the point in the moduli space of the
\ntwo\ theory where the monopole is massless.  Since when
\ntwo\ is broken to \none\ the vacua which survive are those with
massless monopoles or dyons, we will need to discuss in more detail
precisely how this occurs.

\newsec{\none\ Supersymmetry and Confinement: Mesons, Baryons, and 
Strings}
\seclab\nonetheory

In this section we show that in the M theory representation of the
quantum theory of
\none\ supersymmetric gauge theory with heavy quarks, the quarks do 
not exist by themselves.  However, a quark can join onto the MQCD
string identified by Witten \witMa.  Consequently one can show that
quark-antiquark states bound by a string do exist.  One can also show
that baryonic states exist.  In addition one can discuss strings that
carry more than one unit of flux, which are relevant for baryons and
for dynamics of quarks in higher representations than the fundamental.
Our results in this section follow directly from combining the
discussion of the previous section with the results of
\witMa.

\fig{Brane configuration for the \none\ theory. It is obtained from figure
\sutwo\ by rotating the leftmost NS brane from the $v=x^4+ix^5$ plane into
the $w=x^8+ix^9$ plane.  We show the case of $SU(N)$ (there are $N$ D4
branes combined in the central dark line) with two massive flavors
(each given by a semi-infinite D4 brane on the right).}
{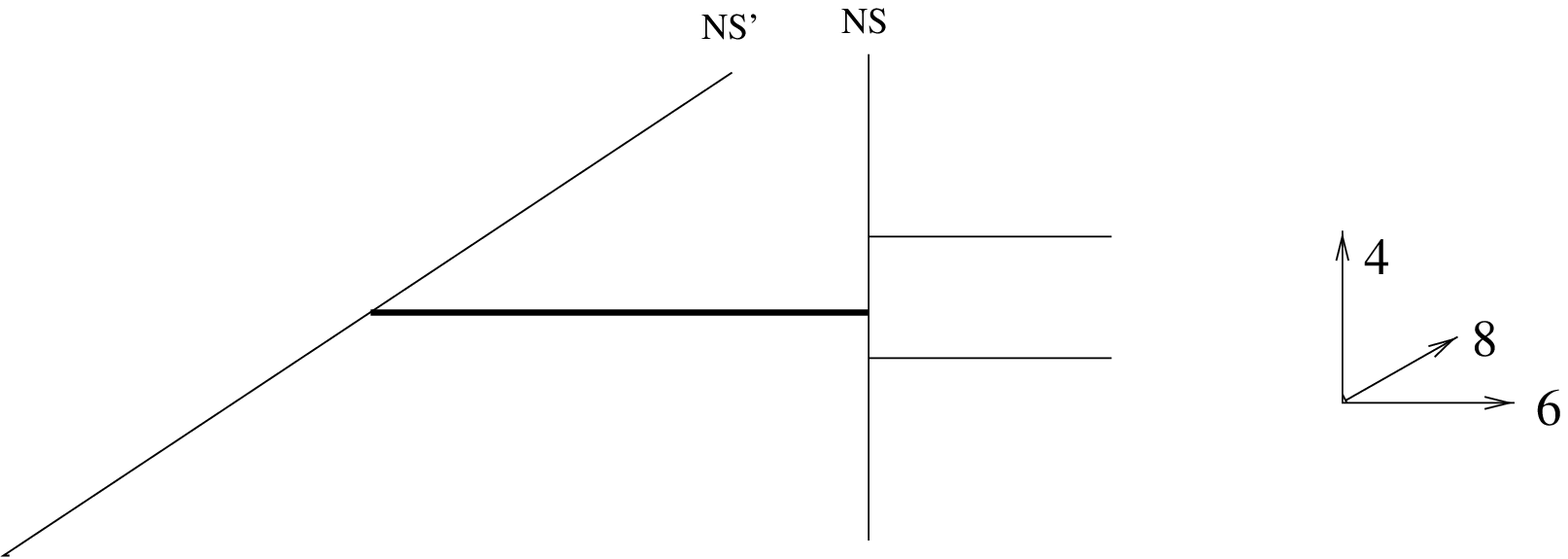}{10 truecm}
\figlabel\massive

The classical construction of the \none\ theory, using the Type IIA string
theory, merely involves rotating the lefthand NS brane of figure
\sutwo\ from the $v=x^4+ix^5$ plane into the $w=x^8+ix^9$ plane \EGK.
This is indicated in figure \massive . (The rotated brane will be
referred to as the NS$'$ brane.) The rotation makes it impossible for
the color branes to move apart, corresponding to the absence of an
adjoint scalar in the \none\ theory.  The flavor branes can still be
placed anywhere in $v$.  Classically, the quarks and gauge bosons are
constructed as Type IIA strings just as in figure \sutwo.

  When the \none\ quantum theory is studied using M theory, the
physics is again described in terms of a Riemann surface $\Sigma$,
which is now embedded in the flat six-dimensional space $v,w,t$, and
is specified by two complex equations in these coordinates.  The curve
for the pure MQCD theory with gauge group $SU(N)$ is given by the
equations
\eqn\pure{vw=\zeta  ; \ v^N=\zeta^{N/2}t;}
this has also been shown in \refs{\witMa,\Berkeley,\yank}.
The constant $\zeta$ essentially determines the MQCD scale
$\Lambda_1$. We will return to the exact relation in the
following.  The addition of two flavor branes at $v=m_1$ and $v=m_2$,
where $m_1,m_2\gg\Lambda_1$, modifies the curve to
\refs{\Berkeley,\yank}
\eqn\puremass{vw={\zeta} \   
; \ v^N=\zeta^{N/2}(1-{v\over m_1})(1-{v\over m_2})t \ .}

\fig{Singularities of $\Sigma$ in the $v$ plane for $SU(6)$ 
with two heavy quarks. The points $m_i$ correspond to positions of the
flavor branes, while $v=0$ corresponds to the position of the
$6$ color branes.  Dashed lines correspond to the intersection of
$\Sigma$ with $x^{10}=0$. The contour $\gamma_0$ wraps $6$ times
around $x^{10}$, while $\gamma_{1,2}$ wrap only once. }
{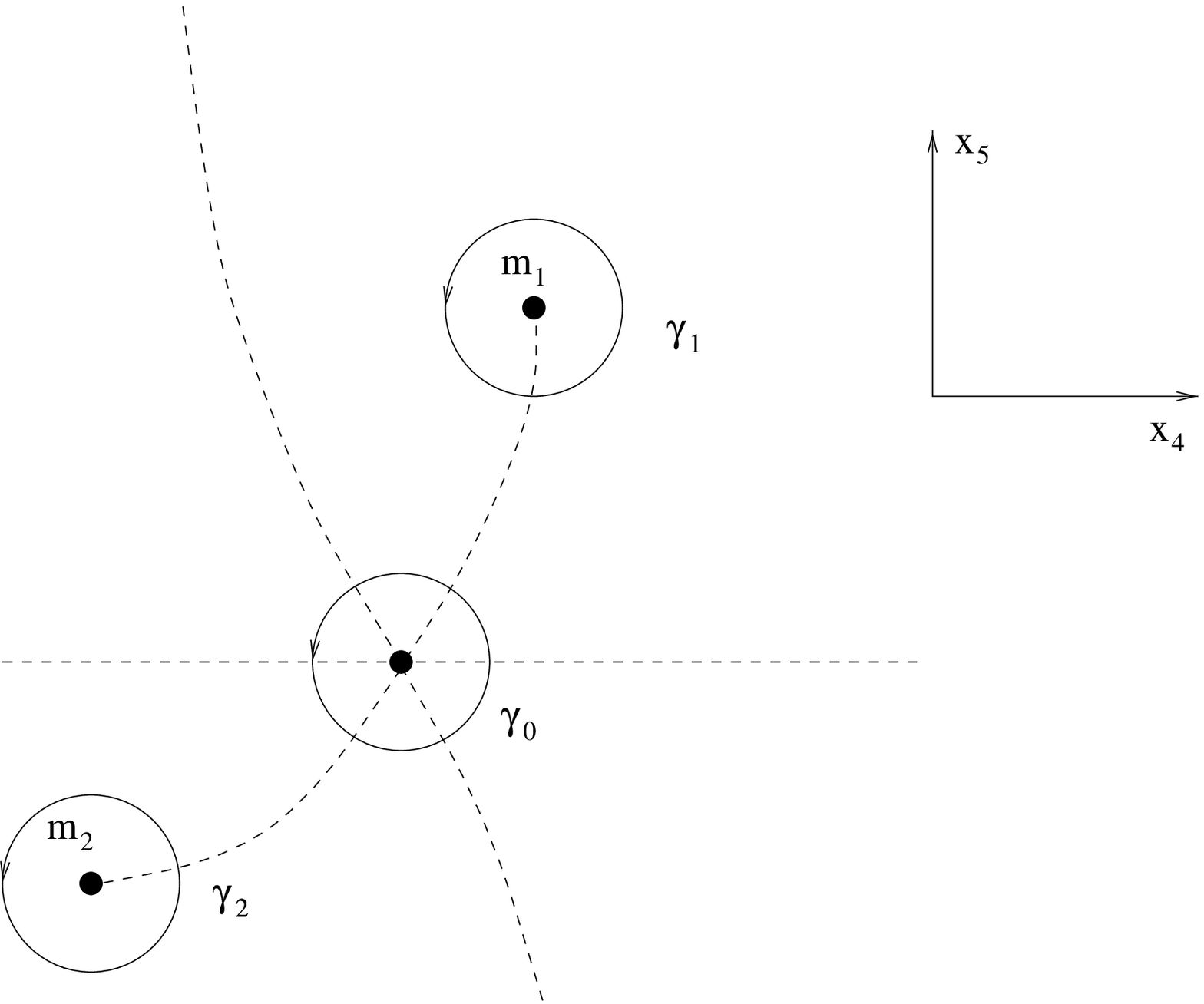}{9truecm}
\figlabel\circles

It is difficult to represent this curve in its entirety because of its
embedding in six dimensions, but we may still consider $w$ and $t$
(now single-sheeted) as a function of $v$.  As shown in figure \circles\
there are three singularities, one at $v=0$ (which corresponds to the
NS$'$ brane) and one each at $v=m_1$ and $v=m_2$.  Note that although
the curves $\gamma_1$ and $\gamma_2$ wrap once around $x^{10}$, the
curve $\gamma_0$ wraps $N$ times around $x^{10}$.

Can we construct gauge bosons or quarks?  As gauge bosons are expected
to be light, they need not be easy to see.  However, the heavy quarks
have such large masses and such a small effect on the dynamics that we
should be able to construct them.  Specifically, we expect
to find states carrying one unit of a flavor quantum number, with mass
of order $m_1$.  In analogy with the
\ntwo\ case above, we should construct such a quark by taking a membrane
with one boundary on $\gamma_1$ and closing it on a curve which wraps
once in $x_{10}$.  However, the only other curve of this type is $\gamma_2$,
which carries a flavor quantum number. Therefore quark states with
one unit of flavor do not exist in the quantum theory.

However, a membrane with a boundary on $\gamma_1$ {\it can} end on a
MQCD string.  To see this, consider in detail a finite two-dimensional
surface given by $\{v,w,t\}(\sigma,\tau)$ ( for $0\leq \sigma,\tau \leq 1$)
with the properties that at each $\tau$ it wraps once around $x^{10}$
\eqn\wrap{
v(\sigma+1,\tau)=v(\sigma,\tau) \ ; w(\sigma+1,\tau)=w(\sigma,\tau) \ ; \
t(\sigma+1,\tau)=e^{2\pi i}t(\sigma,\tau)\ ,
}
and that at $\tau=0$ it intersects the curve $\gamma_1$
\eqn\quark{
v(\sigma,0)=\gamma_1(\sigma)= {\zeta\over w(\sigma,0)} \ ; \
t(\sigma,0)=e^{2\pi i\sigma}t(0,0)\ .  }

\fig{$\Sigma$ (the diagonal lines) is pictured, for
fixed values of $|t|$ and $|v|$, in the $({\rm arg} (v),x^{10})$
plane.  A rotation of $2\pi/N$ in $v$ corresponds to shifting $\Sigma$
once around $x^{10}$. The curve $C_t$, which is closed and passes
through point A, is homotopic to $C_v$, which connects points A
and B on $\Sigma$. The length of $C_t$ is proportional to $R$, while
$C_v$ has length of order $\sqrt{\zeta}$.}  {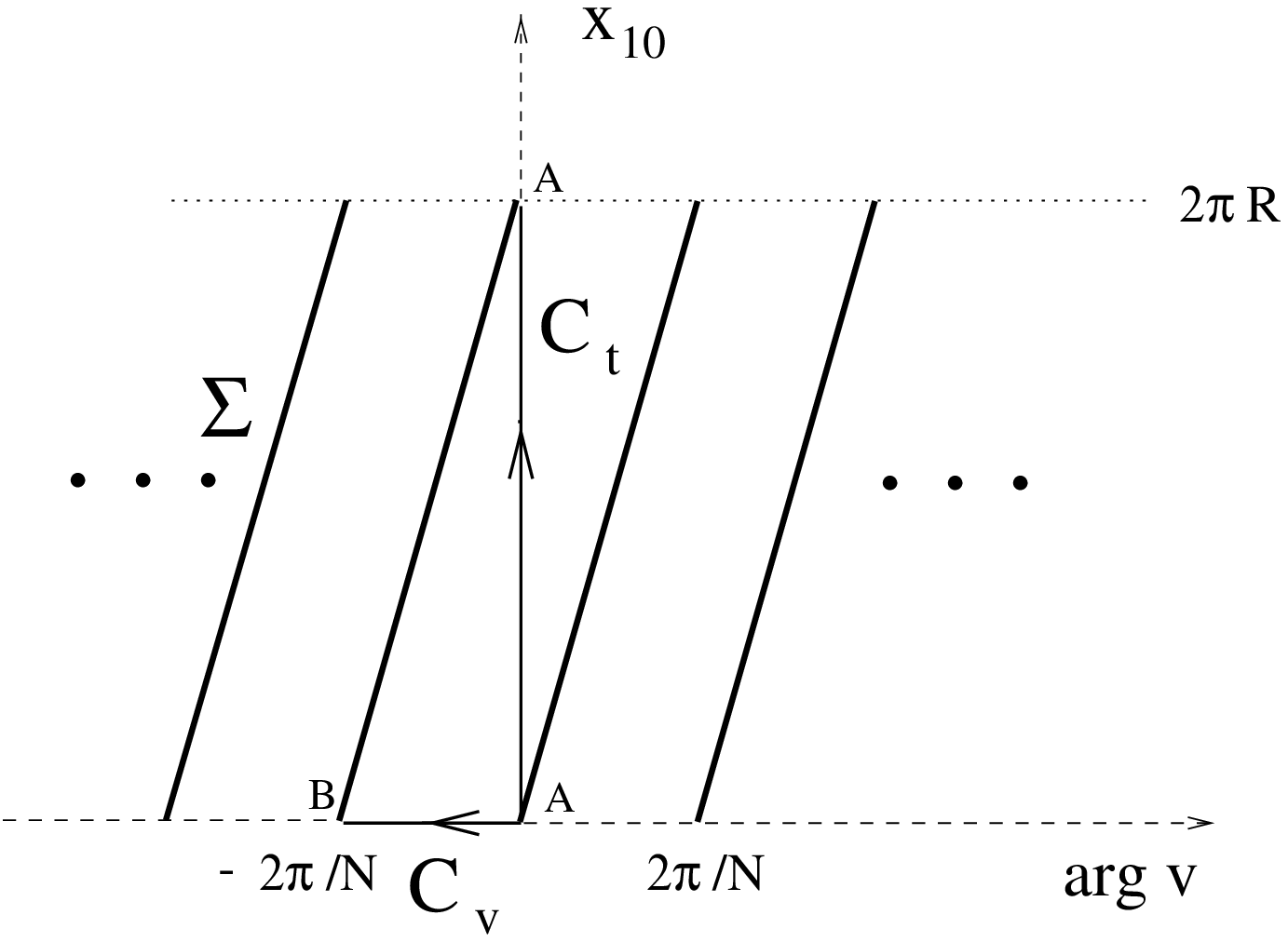}{8truecm}
\figlabel\contour

We may deform this curve (by moving the membrane smoothly inside of
the $v,w,t$ space) so that\foot{In this expression we assume
$\sqrt\zeta t_0^{1/N}\ll m_i$ and approximate $\Sigma$ by its form \pure.}
\eqn\defor{
v(\sigma,1)=\sqrt\zeta t_0^{1/N}= {\zeta\over w(\sigma,1)} \ ; 
\ t(\sigma,1)=e^{2\pi i\sigma}t_0\ . 
 } 
 Then the curve $C_t=\{v,w,t\}(\sigma,1)$
intersects $\Sigma$ (at a unique point A) at $\sigma=0$ and at
$\sigma=1$; see figure \contour\ in which the intersection of $\Sigma$
with $|v|=|t_0^{1/N}|$ is shown along with $C_t$.  But as noted in
\witMa, this curve is homotopic to the curve $C_v$ as is obvious from
the figure.  We emphasize that $C_v$ intersects $\Sigma$ only at
its endpoints and does not represent the end of a membrane.  Instead,
this homotopic transformation represents the opening of a hole in the
membrane world-volume; the curve $C_t$ is closed in $v,w,t$ but $C_v$
is not.

The curves $C_v$ and $C_t$ have quite different physical implications.
A closed membrane wrapping once around the eleventh dimension, like
that of \wrap, is identified in double dimensional reduction with the
elementary Type IIA string, whose tension goes like $R$ in M theory
units.  This string can exist anywhere in ten-dimensional spacetime.
The curve $C_v$, by contrast, has length $\sqrt\zeta$, which is
related to the MQCD scale, and, having boundary on $\Sigma$, gives an
open membrane which must be localized around the fivebrane.  According
to Witten \witMa, the curve $C_v$, when extended into a membrane by
dragging it along a curve $C$ in space, represents a MQCD string lying
on the curve $C$.

 The tension of the MQCD string is proportional to the length
of $C_v$, which is a straight line in $v,w,t$ space 
connecting the points \witMa\
\eqn\points{
A=(\sqrt\zeta t_0^{1/N}, \sqrt\zeta t_0^{-1/N}, t_0) \qquad {\rm and}
\qquad 
B=(\sqrt\zeta t_0^{1/N}e^{-2\pi i/N}, 
\sqrt\zeta t_0^{-1/N}e^{2\pi i/N}, t_0)
}
Its length is
\eqn\lenght{\sqrt{|\Delta v|^2+|\Delta w|^2}=
2\sqrt\zeta \sqrt{t_0^{2/N}+ t_0^{-2/N}}\sin(\pi/N).
}  
To get the MQCD string tension we should further minimize
\lenght\ with respect to $t_0$. Since $\Sigma$ has a symmetry 
(for very heavy quarks) under $t\leftrightarrow 1/t$ which exchanges $w$
and $v$ --- the reflection symmetry that exchanges the
two NS branes --- the minimum will be at $t_0=1$, giving length
\eqn\lenghtdue{2\sqrt{2\zeta}\sin(\pi/N)}
Multiplying by the membrane tension (1 in these units) we get
the MQCD string tension.  If we want to match on to \none\ QCD field
theory expectation, where the string tension should be of order
$\Lambda^2$, we must take $\zeta\sim\Lambda^4$ \witMa.  For the MQCD
string to be stable against decay to Type IIA strings, it must be that
its tension $\sim\sqrt\zeta$ is much less than $R$
\witMa.\foot{Actually the minimal tension is given not by $C_v$ but by
the nearby line which intersects $\Sigma$ at right angles; the
difference in lengths is very small, of order $\sqrt\zeta/R$, and can
be ignored until section 9.}

\fig{An abstract rendering of the brane which represents the meson,
showing its topological properties.  The three boundaries where it
intersects $\Sigma$ are highlighted.  Its projection into spacetime is
shown.}{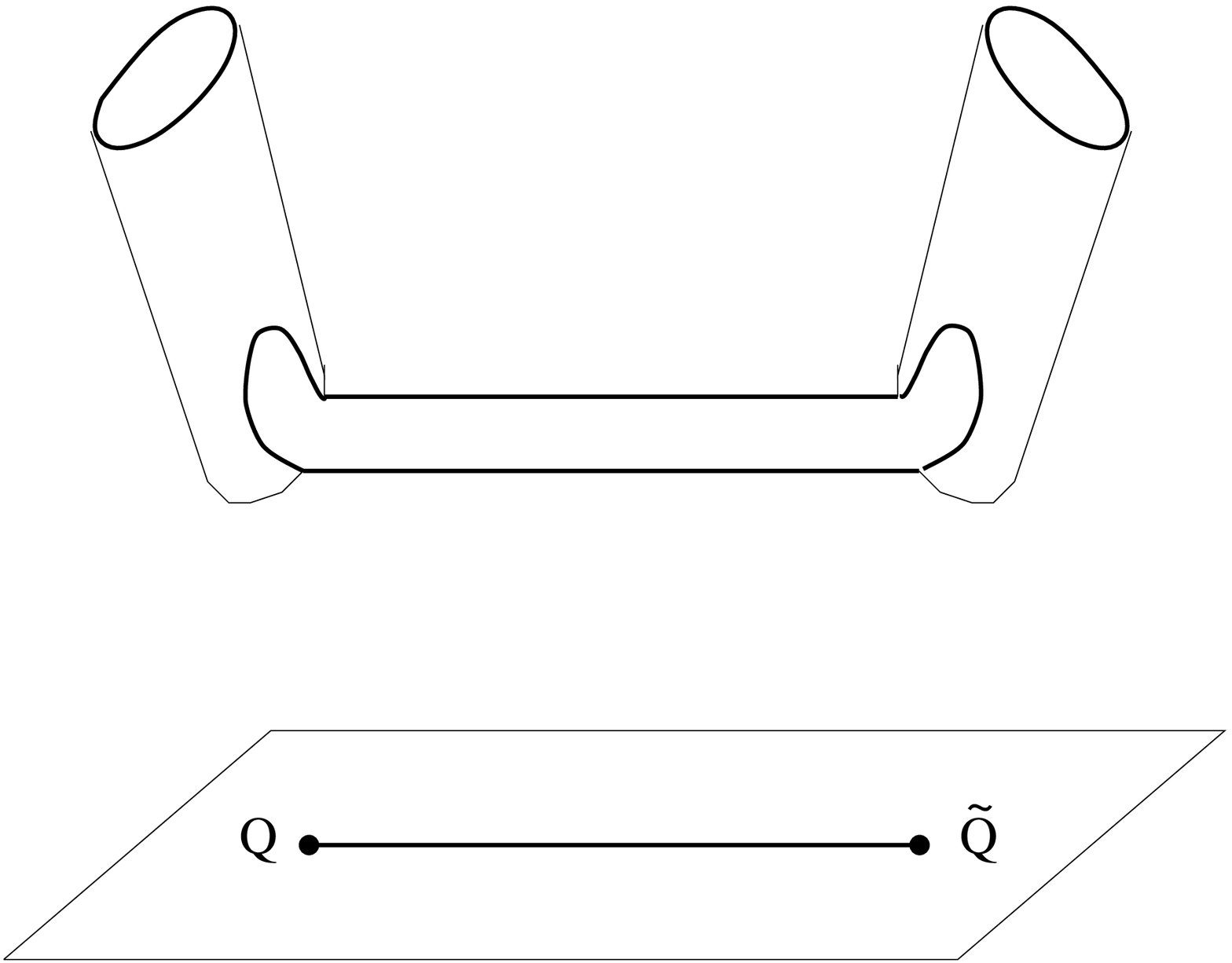}{10 truecm}
\figlabel\mesonbrane

Thus, a quark can connect to a MQCD string, and we can form a meson by
connecting the string to an antiquark at the other end.  A meson
$Q(x)$---$\tilde Q(y)$ is thus given by a single membrane with three
boundaries on $\Sigma$, one on $\gamma_1$ at the point $x$, another on
$\gamma_2$ at the point $y$, and a third which opens up at $x$ during
the transition from $C_t$ to $C_v$, stretches along the MQCD string
$C$, and closes again via the reverse homotopy at $y$ (figure
\mesonbrane).  This picture is quite similar to that given for abelian
confinement by Greene, Morrison and Vafa \gmv\ though it has
significant differences also.  The mass of the meson will be roughly
given by the sum of the quark masses and the tension of the MQCD
string times its length, in agreement with expectations, as long as
the meson is neither too long nor too short.  For short strings a
Coulomb potential, rather than a linear potential, applies between the
quarks, but this effect is not visible in the semiclassical membrane
picture of a meson.  Long strings can break via heavy quark pair
production; this process can easily be seen in the membrane picture,
though we omit any further discussion here.\foot{ It naively appears
that the mesons we have identified can decay to membranes stretching
directly from $\gamma_1$ to $\gamma_2$, namely to gauge bosons of the
flavor group.  We have explained in section \twotheory\ why the flavor
gauge bosons do not couple dynamically to the four dimensional
theory.}

\fig{The homotopy transformation by which $5$ quarks form a baryon of $SU(5)$.
The notation is as in figure \contour, except that here we emphasize
the periodicity in $\arg v$. The vertical curves $C_t(i)$ can be
homotopically deformed to the horizontal curves $C_v(i)$, which can be
joined together in a closed loop that can be detached from $\Sigma$.
} {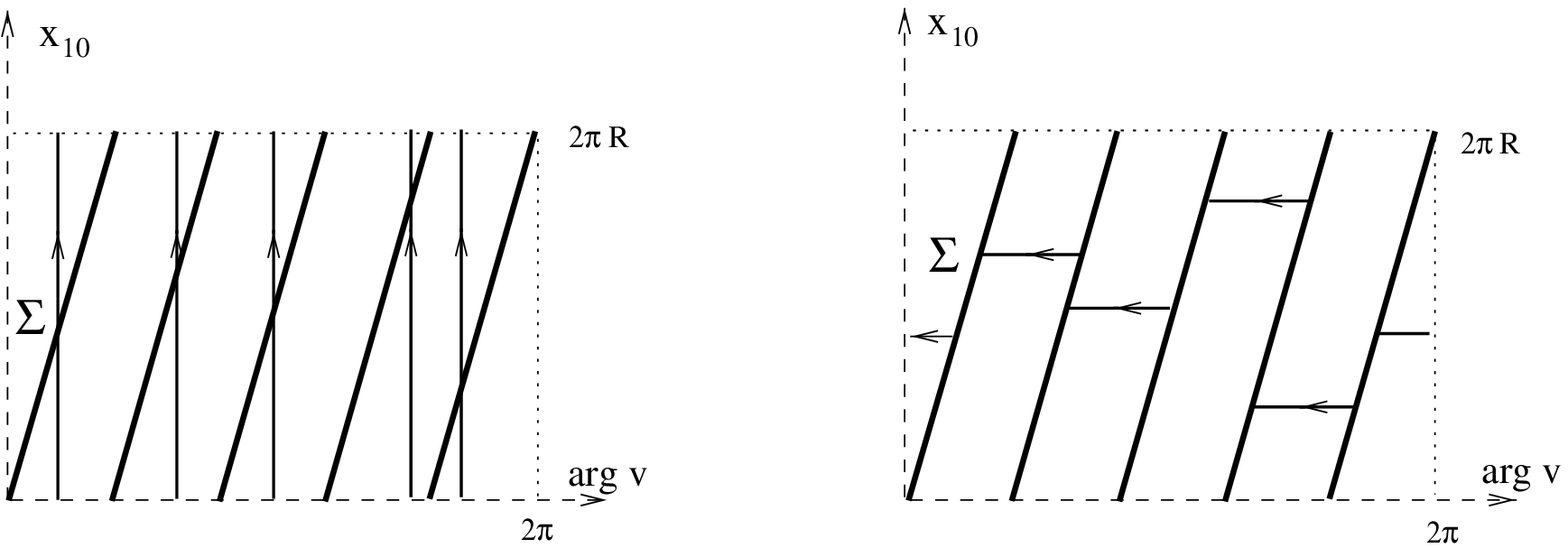}{14 truecm}
\figlabel\baryarg

Next, we construct a baryon from $N$ massive quarks.  It is convenient
to add many flavor D4 branes to the theory, each one giving a
singularity in $v$ around which $x^{10}$ winds once.  We can attach
$N$ membranes to $N$ contours surrounding these singularities, and
bring them toward the origin in $v$.  Following \witMa, and as shown
in figure \baryarg, the quarks can be brought to the curves $C_t(i)$
and then homotopically deformed to the curves $C_v(i)$; the latter can
be joined together into a closed loop that can then be pulled off of
the surface $\Sigma$, following which it can be shrunk to a point in
the $v,w,t$ space.  This means that a baryon consists of a single
membrane with $N$ boundaries, each wrapping (with the same
orientation) around one of the singularities at $v=m_i$, along with a
single additional boundary running along the $N$ strings and ending at
the vertex which joins them together.

As a final comment, we note that when some number $N_f<N_c$ quarks are
taken to be light compared with the MQCD scale, the situation is
topologically the same but requires physical reinterpretation.  (A
different phase structure emerges for $N_f\geq N_c$.)  The $N_f$ light
squarks $q_r$ acquire expectation values, breaking $SU(N)$ to
$SU(N-N_f)$; their components are eaten by gauge bosons, except for
$N_f^2$ light singlets.  The heavy quarks $Q^i$ split into components
which are charged under $SU(N-N_f)$ and components which are not.  It
is easy to see that the mesonic membranes connecting two light quarks
make up the $N_f^2$ light singlets, those connecting a light and a
heavy quark are the neutral components of the heavy quarks, while
those connecting two heavy quarks are quark-antiquark mesons confined
by the $SU(N-N_f)$ interaction.  This conforms with field theory
expectations.

\newsec{Strings with Multiple Units of Flux}
\seclab\multiple

One expects, in general, as many as $N-1$ stable $k$-strings in
ordinary QCD, characterized by their quantum number $k$ under the
center of the gauge group, or equivalently by the number of units of
flux $k$ which they carry (mod $N$).  The tension as a function of $k$
is both periodic in $N$ and symmetric under $k\sim N-k$, so the
strings can have as many as $\integer{N/2}$ different tensions.  The
stability of the $k$-strings with $k>1$ is of physical interest.  For
example, a quark in the antisymmetric representation may either
connect to two $1$-strings or to a single $2$-string.  Since the two
choices are not distinguished by a quantum number, the preferred
configuration is determined dynamically.

\fig{The curve $\Sigma$ is shown for $SU(6)$, with the circle $|v|=\sqrt\zeta$ 
highlighted.  Intersections of $\Sigma$ with $x^{10}=0$ are shown with
dashed lines. The line AB, which lies outside $\Sigma$ except at its
endpoints and is at a {\it fixed} value of $x^{10}$, corresponds to a
string with one unit of flux. It is clear that the line AC (a string
with two units of flux) is energetically preferred to the line ABC
(two strings with one unit of flux each.)}  {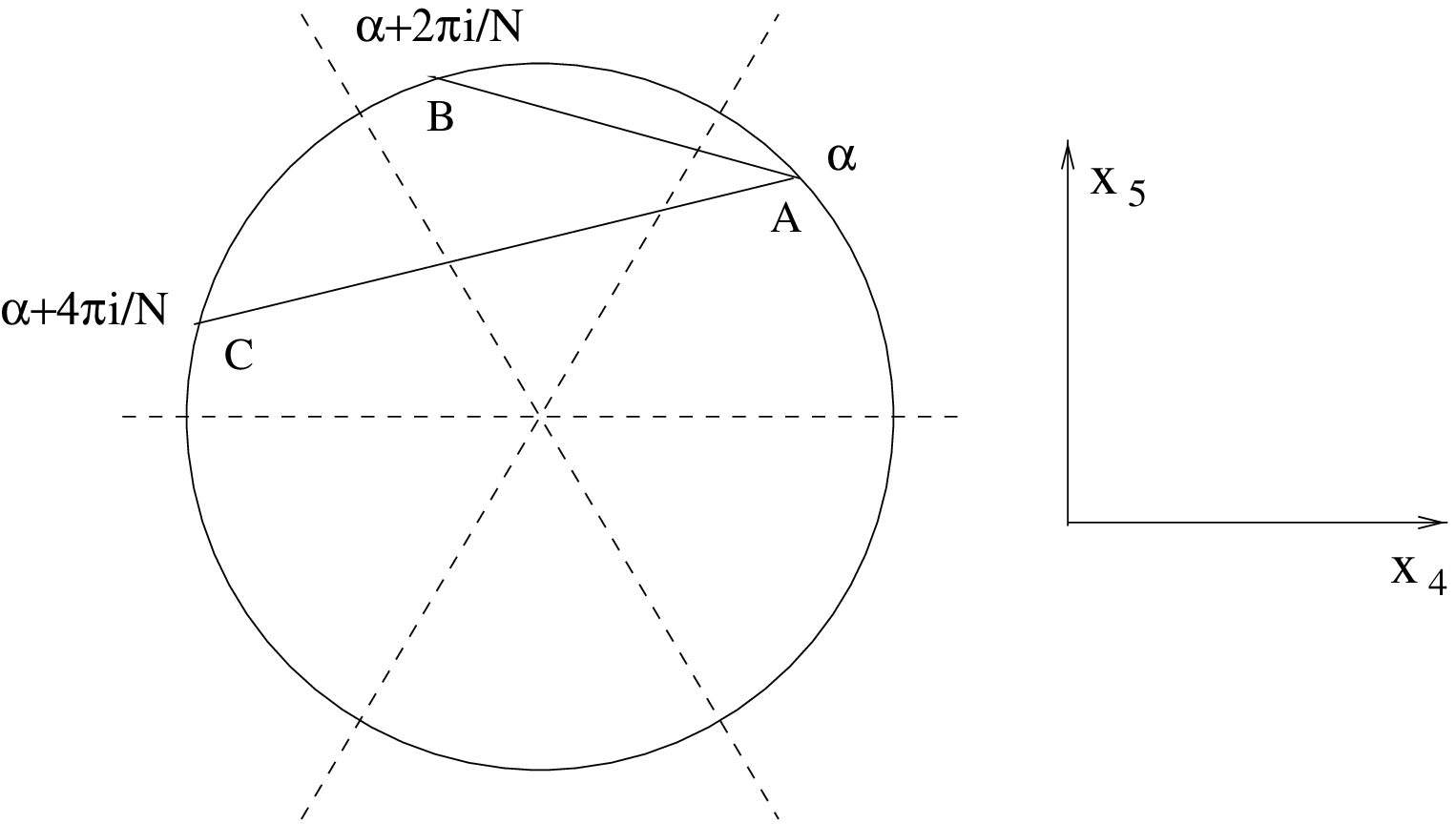}{10 truecm}
\figlabel\circolo

In MQCD, the $2$-string is preferred.  The tension
of a $k$-string is given by considering the minimal tension of a
string which can connect to $k$ quarks.  A membrane connected to $k$
quarks wraps $k$ times around $x^{10}$.  Using the homotopy
transformation of figure \contour, such a membrane can be rotated into
$k$ MQCD strings of one unit of flux, each of which connects two
points $v_0$ and $v_0e^{2\pi i/N}$ on $\Sigma$, or by a single
$k$-string given by connecting two points $v_0$ and $v_0e^{2\pi i
k/N}$ (figure \circolo).  It is obvious that the latter possibility
leads to the shorter curve, whose length, as in \points-\lenghtdue, is
given by
\eqn\klengths{
2\sqrt{2\zeta}\sin {\pi k\over N},
}
so the $k$-string tensions are proportional to $\sin {\pi k\over N}$.
The importance of this formula and its relevance for ordinary QCD will
be discussed in the coming sections.

\newsec{Breaking \ntwo\ Supersymmetry to \none}
\seclab\softly

The images of strings, mesons and baryons from the previous sections
are in contrast to those which emerge when pure \ntwo\ $SU(N)$
Yang-Mills theory is weakly broken, at least for $N>2$.
In this section we review the results of Douglas and Shenker \DS, who
analyzed this breaking in detail, generalizing the approach of
\swone.  We show that M theory reproduces these results, and discuss 
the picture that it suggests for the transition from the physics of
weakly broken \ntwo\ gauge theory to that of pure \none\ gauge theory.

\subsec{Brief Review of Weakly Broken Pure \ntwo\ Gauge Theory}
\subseclab\review
 
The \ntwo\ vector multiplet consists of an \none\ vector multiplet
along with a chiral multiplet $\phi$ in the adjoint
representation. With addition of a mass term $W=\mu u$, where
$u=\half\tr\phi^2$, \ntwo\ \susy\ is broken to \none.

In the quantum \ntwo\ $SU(N)$ gauge theory \refs{\swone\sun}, the low
energy effective theory on the Coulomb branch has gauge symmetry
$U(1)^{N-1}$. The elliptic curve for this theory has been studied by
various authors \refs{\sun,\suntwo,\sunthree,\sunfour}, and, as
discussed for $SU(2)$ in section \twotheory, can be identified as part
of the world-volume of a fivebrane \witMb. The curve $\Sigma$ is given
by
\eqn\curv{t + \Lambda_2^{-2N}P_N(v) +1/t=0,}
where $P_N(v)$ is a polynomial of order $N$ and $\Lambda_2$ is the
dynamical scale of the \ntwo\ theory. The curve has the manifest
symmetry $t\leftrightarrow 1/t$.

  The theory has $N$ special vacua at which $N-1$ mutually local
monopoles or dyons become massless \swone.  Only these vacua survive
when \ntwo\ \susy\ is broken.  These vacua are related by a symmetry,
so without loss of generality we limit ourselves to the one with
monopole states.

In the monopole variables, the mechanism of \ntwo\ breaking very
closely resembles the addition of Fayet-Iliopolous terms to $N-1$
decoupled \ntwo\ \susic\ Abelian Higgs models.  One can choose a basis
for the $N-1$ $U(1)$ factors in which each monopole is charged under
only one magnetic photon \refs{\sun,\suntwo,\DS}; that is, their magnetic
charges in this basis are $(1,0,0,\cdots,0)$, $(0,1,0,\cdots,0)$,
$(0,0,1,\cdots,0),\dots$, $(0,0,0,\cdots,1)$.  The superpotential for
the monopoles is then
\eqn\superpot{
W = \sum_{p=1}^{N-1} \sqrt{2} a_D^{(p)}M_p\tilde M_p + \mu u(a_D) } 
where $a_D^{(p)}$ is the scalar in the vector multiplet of the $p$-th
$U(1)$ factor, $(M_p,\tilde M_p)$ is the $p$-th monopole
hypermultiplet, and 
\eqn\uformula{
u(a_D) = b\Lambda_2^2 - \sum_j c_j \Lambda_2
a_D^{(j)}+ {\cal O}(a_D^{(i)}a_D^{(j)}) \ ,} 
where $b,c_j$ are constants determined by the solution of the theory
\refs{\sun,\suntwo,\DS}.  The linear term in $a_D^{(j)}$ would be an
(\ntwo)--preserving Fayet-Iliopolous term for the $j$-th $U(1)$
factor, and the analysis would be identical to that of the Abelian
Higgs model, were it not for the higher order terms in $u(a_D)$ which break
\ntwo\ \susy.  The potential energy is minimized for 
$\vev{M_p\tilde M_p}=c_p \mu\Lambda_2$ and $a_D^{p}=0$.  In each
$U(1)$ factor, the non-zero monopole expectation value breaks the
gauge symmetry and permits a Nielsen-Olesen string solution to the
classical equations \NielOle.
 One finds \DS\ that the $N-1$ strings have
string tensions
\eqn\tens{
T(p) =2\pi |\vev{M_p\tilde M_p}| 
= 4\sqrt{2}\pi  |\mu\Lambda_2| \sin{\pi p\over N}
 } 
where $p$ runs from $1$ to $N-1$.  Note the symmetry under
$p\leftrightarrow N-p$.  The calculation is reliable for small $\mu$
since the monopole Lagrangian is weakly coupled.  (The failure of
these strings to be BPS saturated will be discussed in section 7.)

To understand how these strings manifest themselves physically, it is
essential to note the following.  Since monopoles have condensed, the
flux stemming from electrically charged states must be confined into
strings.  In the basis mentioned above, in which the monopole
$U(1)^{N-1}$ magnetic charges are simple, a heavy quark in the
fundamental representation has $N$ color states $Q^1,Q^2,Q^3,\cdots
Q^N$, with electric charges $(1,0,0,\cdots,0)$, $(-1,1,0,\cdots,0)$,
$(0,-1,1,\cdots,0),\dots$, $(0,0,0,\cdots,-1)$.  From this it is easy
to see that the strings of Douglas and Shenker are indeed
distinguished by the amount of flux that they carry.  The state
$(Q^1Q^2Q^3\cdots Q^p)$ has charge $p$ under the center of the group
and must couple to strings carrying a total of $p$ units of
flux. Since this state is charged under the $p$-th $U(1)$ factor and
neutral under the others, it couples only to the string labelled $p$
above, and so the $p$-th string is indeed a $p$-string.  

In addition, the charges of the state $Q^p$, which carries one unit of
flux, are such that it must attach to two strings, a $p$-string and a
$(p-1)$-antistring.  Since this is the case, we are led to the
surprising conclusion that there are actually many types of heavy
quark mesons \DS.  A (highly excited) $Q^p$---$\tilde Q_p$ state will
be bound by a string of tension $T(p)$ and an antistring of tension
$T(p-1)$, so there are, in all, $\integer{(N+1)/2}$ distinguishable
sets of meson states made from a quark and antiquark in the
fundamental representation.

\fig{Standard picture for baryons in $SU(6)$: six quarks each with
a $1$-string connected by a six-string vertex.}{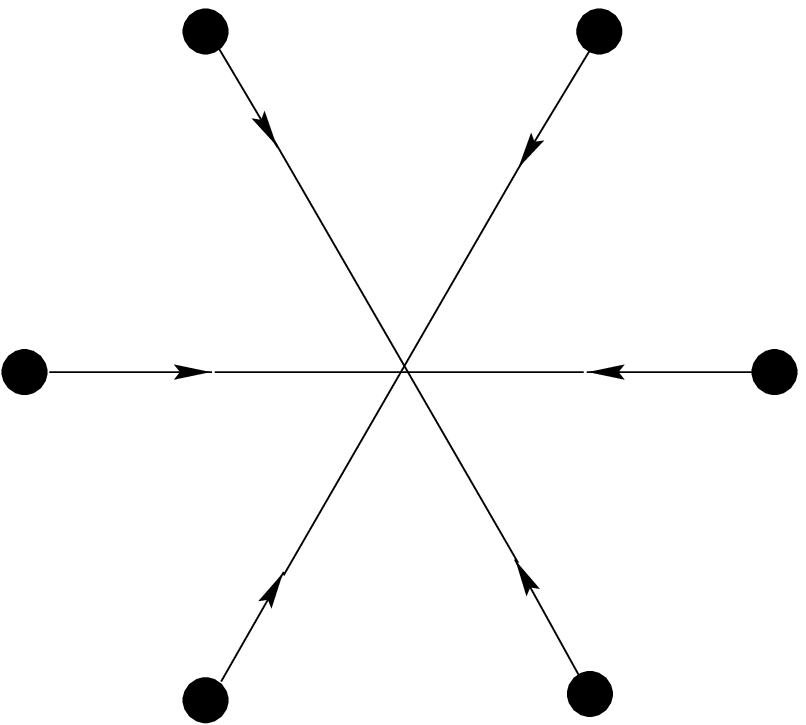}{5 truecm}
\figlabel\baryons
\fig{Expectation for a baryon in weakly broken \ntwo\ $SU(6)$:
the $k$-th quark is connected to a $(k-1)$-string on the left and a
$k$-string on the right.}  {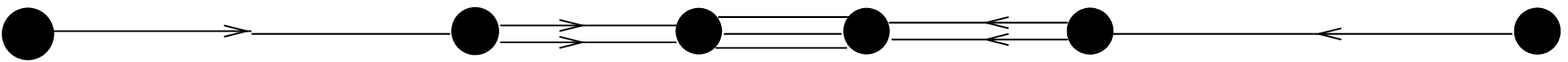}{10 truecm}
\figlabel\dsbaryons

These features also affect the baryons.
A common picture for a baryon in QCD is that of figure
\baryons.  Here, the expectation would more naturally be that of
figure \dsbaryons\ \DS.

Note that a sufficiently excited state $Q^p$---$\tilde Q_p$ is only
metastable, due to pair production of $W$ bosons \DS.  If an excited
state with length $L$ is long enough that the quantity
$[T(p)+T(p-1)-T(1)]L$ is greater than twice the mass of the $W$ boson
$[W_{\alpha}]_p^1$, then the state $(Q^p [W_{\alpha}]_p^1)-
([W^{\alpha}]_1^p\tilde Q_p)$ will have lower energy (for the same
angular momentum) than the $Q^p-\tilde Q_p$ state. However, one can
always find a range for $L$ in which the $Q^p-\tilde Q_p$ states are
metastable and can be observed.

This theory differs significantly from the expectation for
\none\ or non-supersymmetric QCD, where only one type of meson is
anticipated.  The key point \DS\ is that in weakly broken \ntwo\ gauge
theory, the Weyl group of $SU(N)$ is spontaneously broken by the
expectation value of the field $\phi$.  This makes it possible for the
$Q^p-\tilde Q_p$ bound state spectra to depend on $p$.  The absence of
scalar fields in \none\ and in non-\susic\ QCD makes it plausible that
the Weyl group is unbroken in these theories.

However, the $k$-strings of weakly broken \ntwo\ QCD and those of
\none\ MQCD share an important property: although
their tensions \tens\ and \klengths\ have different overall
normalizations, they both satisfy the formula
\eqn\tensionratio{
{T(k)\over T(k')} = {\sin {\pi k\over N}\over \sin {\pi k'\over N}} \
.  }  
(We will see in section 9 and appendix B that this formula even
applies for nonsupersymmetric MQCD.)  Still, the dynamics of the
Douglas-Shenker and \none\ MQCD strings are somewhat different, as the
MQCD picture makes clear.  We should also explore the physics behind
the disappearance of the $\integer{(N+1)/2}$ mesons in favor of the
single meson of \none\ MQCD.  In the remainder of this section we will
discuss the transition from small to large $\mu$ in detail.

\subsec{The M Theory Fivebrane for Broken \ntwo\ Gauge Theory}

\fig{The curve $\Sigma$ for \ntwo\ $SU(6)$ gauge theory
near the massless monopole point.  The masses
of the monopoles, proportional to the area of the holes in the
surface, are all becoming small.}
{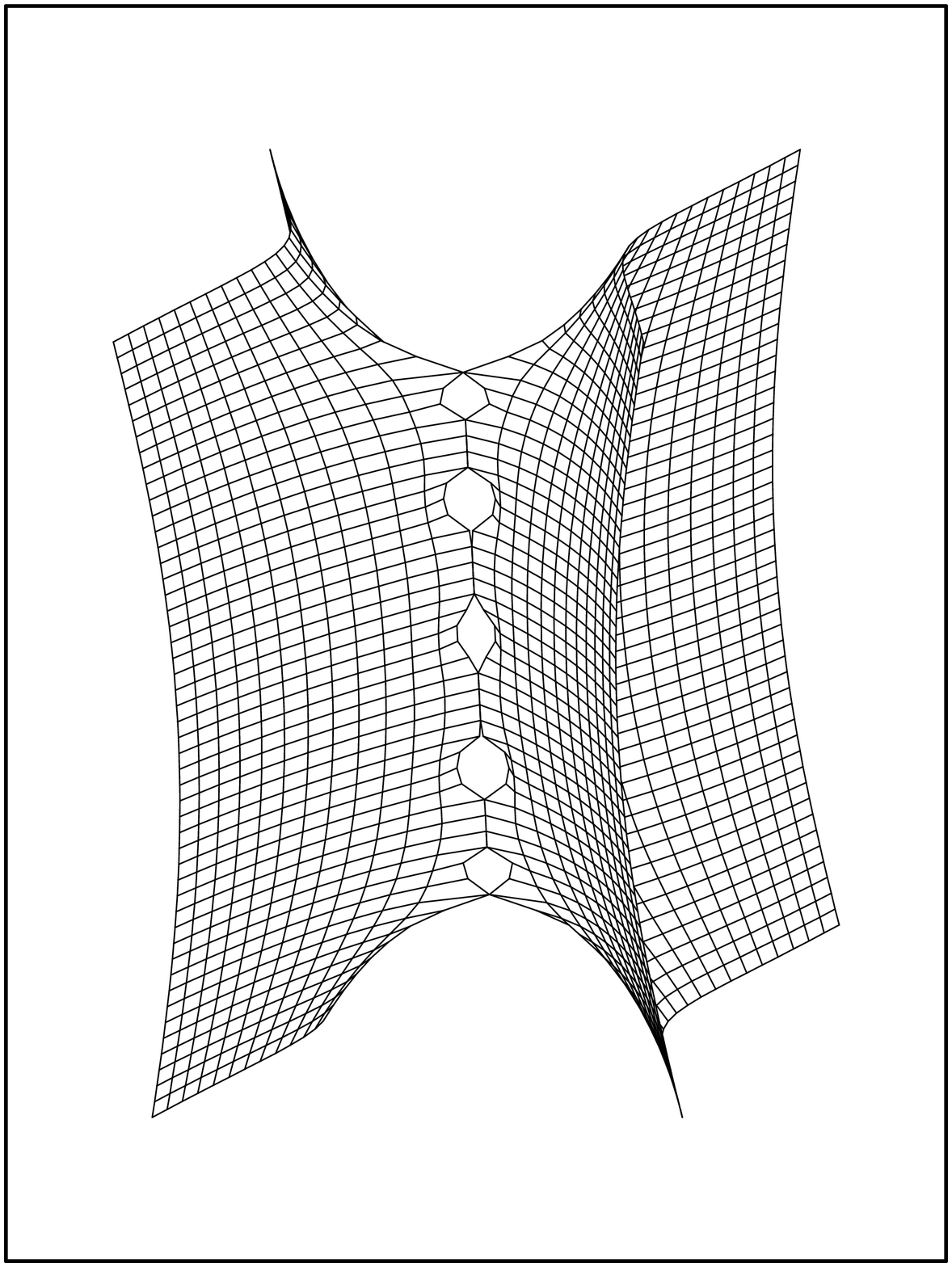}{8truecm}
\figlabel\lightmon

 We now examine the fivebrane theory as a function of $\mu$.  As
a starting point, we identify the properties of the fivebrane
describing the \ntwo\ theory as we approach the vacuum with massless
monopoles.  In this limit the areas of the holes in the genus $N-1$
Riemann surface simultaneously shrink to zero, as seen in figure
\lightmon; the monopoles become massless, and
the curve $\Sigma$ degenerates to a genus zero surface.

\fig{The curve $\Sigma$, projected into the $v$ plane, for $SU(6)$ 
at the massless monopole point.  The two halves of $\Sigma$ meet at
$|t|=1$ along a line.  The line segments are the colored D4 branes.
Massless monopoles are localized at the positions of the filled
circles; the open circles are ends of D4 branes where there are no
monopoles. We indicate $t$ real and positive (negative) using dashed
(dotted) lines. In going between two dashed lines or two dotted lines,
$\Sigma$ wraps once around $x^{10}$.}  {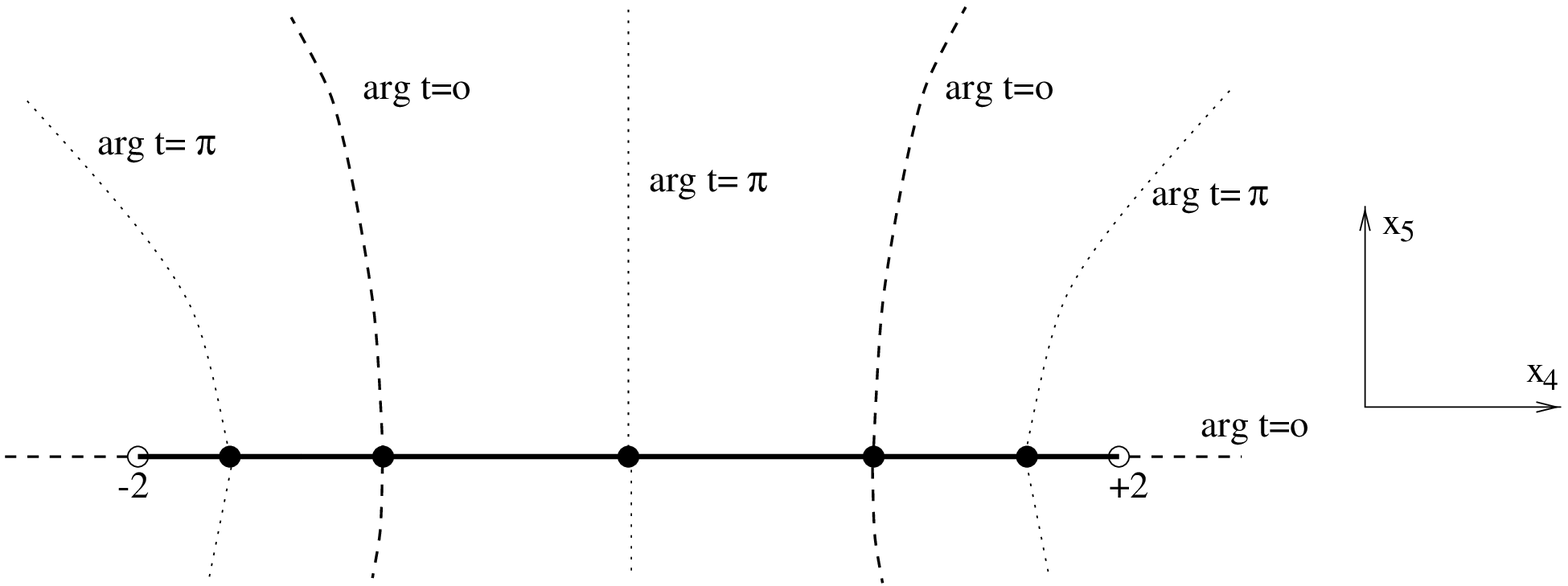}{13truecm}
\figlabel\lineseg

The equation for the degenerated surface is \DS\foot{Here we need to be
more careful with units.  We use those appropriate for M theory, and
put $l_P=1$.  Additional details are given in appendix A.}
\eqn\deg{t = e^{iN\sigma};\quad
2\pi Rv = \Lambda_2\left (t^{1\over N} + t^{-{1\over N}}\right ) =
2\Lambda_2\cos\sigma }
where $\sigma$ is complex.  The two
halves of $\Sigma$ join where $\sigma$ is real, on the line shown in
figure \lineseg. The $N$ line segments touch at $t=\pm 1$,
that is, at $v=2(\Lambda_2/2\pi R)\cos (\pi k/N)$.  The $N-1$
massless monopoles are localized at these points on the
fivebrane.  Note that the $W$ bosons (given by membranes wrapping
around two different line segments) and heavy quarks (given by
membranes wrapping at one end around one line segment and at the other
around a flavor brane singularity at $v=m$) remain massive despite
the degeneration. The presence of $2\pi R$ in equation \deg\ is simply
explained by checking the mass for a $W$ boson. The $W$ boson is a
membrane wrapped around $x^{10}$ and with a length along the line of
order $\Lambda_2/2\pi R$. Its mass is therefore proportional to
$\Lambda_2$, as expected.

What happens when the \ntwo\ theory is broken by the mass $\mu$ for
the adjoint chiral superfield $\phi$?  The NS branes are rotated
relative to one another \refs{\EGK,\barb}, and the surface becomes
\refs{\Berkeley,\witMa}
\eqn\bip{2\pi Rv=\Lambda_2(t^{1\over N} + t^{-{1\over N}}),
\qquad 2\pi Rw=\alpha\Lambda_2t^{1\over N} } 
We can roughly identify $|\alpha|$ with the tangent of the rotation
angle, and $\mu$ with $\alpha/2\pi R$. We are not careful with overall
normalizations. For $\alpha =0$ this clearly reduces to the the \ntwo\
curve since $w$ goes to zero.

For any value of $\mu$ the surface is still symmetric under the
reflection which exchanges the two NS branes (the regions where $v$ or
$w$ go to infinity).  The symmetric point where we expect the surface
to have minimal size is
\eqn\tttt{|t|^{1/N}= t_0^{1/N} = 
\left |{1\over (1+|\mu R|^2)^{1/4}}\right | .}
The intersection of $\Sigma$ with this hyperplane is therefore the curve
\eqn\curva{
t=t_0e^{2\pi iN\sigma} ;
v  = {\Lambda_2\over 2\pi R}\left 
  ( (1+|\mu R|^2)^{1/4}e^{-2\pi i\sigma}+ {e^{2\pi i\sigma}
     \over (1+|\mu R|^2)^{1/4}}\right ); 
w= {\Lambda_2\mu e^{2\pi i\sigma}\over(1+|\mu R|^2)^{1/4}}e^{2\pi i\sigma}
}
which describes an ellipse in the $v$ plane with semi-axes 
\eqn\semiax{{\Lambda_2\over 2\pi R}((1+|\mu R|^2)^{1/4}\pm (1+|\mu 
R|^2)^{-1/4})}
and a circle in the $w$ plane with radius ${\Lambda_2\mu\over(1+|\mu
R|^2)^{1/4}}$. 

In the \ntwo\ limit $\alpha\rightarrow 0$, the circle in the $w$
plane, which has radius proportional to $\alpha$, shrinks to zero size
while the ellipse in the $v$ plane shrinks to a line, reproducing
equation \deg.  

In the \none\ limit $\alpha\rightarrow \infty$, the ellipse \curva\
becomes a circle, as in figure \circolo.  To see this requires a
rescaling of variables \Berkeley.  The curve \bip\ has a smooth limit
provided that we rescale $t$ as follows\foot{This is a simple
translation in $x^6$ which does not affect the physics of the system.}
\eqn\resc{t^{1/N}={\tilde t^{1/N}\over \sqrt{\alpha}} \ .}  
Matching the \none\ and \ntwo\  QCD scales $\Lambda_1$  and $\Lambda_2$,
using
\eqn\scales{\Lambda_1^{3N}=\mu^N\Lambda_2^{2N}\ ,}
we find in the limit the \none\ curve
\eqn\recall{\sqrt{2\pi R}v=\Lambda_1^{3/2}\tilde t^{-1/N},
\qquad \sqrt{2\pi R}w=\Lambda_1^{3/2}\tilde t^{1/N}\ .}
In the notations of section \nonetheory, $\zeta=\Lambda_1^3/2\pi R$. As
shown in \witMa\ and reviewed in appendix A, we must fix $R\sim
1/\Lambda_1$, in order that $\zeta\sim\Lambda_1^4$ and the string
tension, from equation \lenghtdue, be proportional to $\Lambda_1^2$.

\fig{The ellipse \curva\ in $\Sigma$ for the weakly broken \ntwo\ $SU(6)$
theory. The dashed (dotted) lines correspond to the intersection of
$\Sigma$ with $\arg t=0$ ($\arg t=\pi$).}  {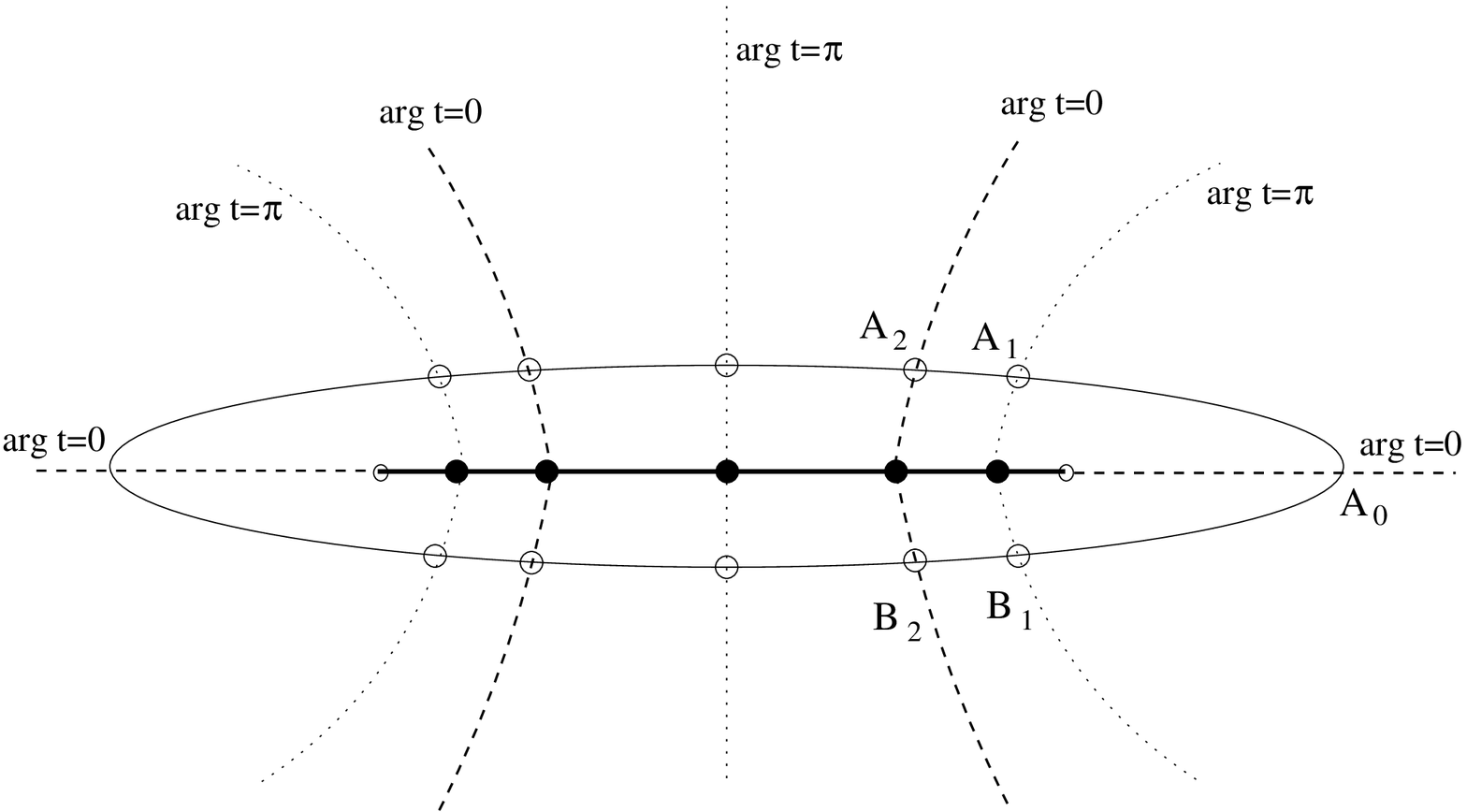}{12truecm}
\figlabel\ellipse

Figure \ellipse\ focuses on the relevant region of $\Sigma$ for the
weakly broken \ntwo\ theory; the ellipse is the curve \curva.  The
dashed (dotted) lines correspond to the intersection with $\arg t=0$
($\arg t=\pi$). Traveling from one dashed line to the next inside
$\Sigma$ involves wrapping once around $x^{10}$.  The points $A_k$ and
$B_k$ are important in our later discussion.  They are given by
$t=(-1)^k$.  Notice that as $\mu \rightarrow 0$ they intersect at the
double branch points where the massless monopoles were localized for
$\mu=0$.

\subsec{Physics of the Transition from \ntwo\ to \none\ Gauge Theory}
\subseclab\physics

 We will now discuss how strings and mesons behave for large and small
$\mu/\Lambda_2$, using MQCD as a guide.  In
particular we will see how the $\integer{(N+1)/2}$ mesons of weakly
broken \ntwo\ go over to a single stable meson in the
\none\ limit.  We will also study the modification of the $N-1$
strings during the transition, and explain why \tensionratio\
holds for all values of $\mu$.

First, we confirm  that confinement of quarks does indeed 
occur when $\mu\neq 0$.  As in the pure
\none\ case, this follows from the fact   
that when $\mu$ is non-zero there are no closed curves in $\Sigma$
(except for those corresponding to flavor branes) which wrap once
around $x^{10}$.  From the equation \bip\ for $\Sigma$, 
it can be seen that any closed curve in $v$ with this property 
cannot close in the variable $w$ for $\mu\neq 0$.  We conclude
there are no heavy quarks.

However, Nielsen-Olesen strings do exist, and for small $\mu$ their
properties agree with the results of Douglas and Shenker, as we will
now show.  As discussed in section \multiple, a string carrying $k$
units of flux (a $k$-string) in pure \none\ MQCD is specified by a
finite real curve in $v,w,t$ with the following properties.  It should not
wrap around $x^{10}$, so that its length is much less than $R$; its
endpoints should lie on $\Sigma$, so that it gives an open membrane;
and any curve {\it inside} $\Sigma$ joining its endpoints should wrap
$k$ times around $x^{10}$, so that this string can be attached to a
state containing $k$ quarks.  Loosely, in the language of figure
\ellipse, between the two endpoints there must lie $k$ dashed lines in
$\Sigma$.

For the weakly broken \ntwo\ theory, we must determine the curves of
minimal length which satisfy these conditions.  By the symmetry which
exchanges the two NS branes, the endpoints of these curves must lie on
the ellipse \curva.  For $k=1$, we need two points on \curva\ which
are separated by one wrapping in $x^{10}$.  It is clear from figure
\ellipse\ that the straight line connecting $A_1$ and $B_1$ 
is the shortest line available.  (The line connecting $A_{N-1}$ to
$B_{N-1}$ would also do.)  For $\mu$ small the major axis of the
ellipse is of order $\Lambda_2/R$ and the minor axis of order
$\Lambda_2\mu^2 R$, while the radius of the $w$ circle is of order
$\Lambda_2\mu$. The length of the straight line between $A_1$ and
$B_1$, $\sqrt{|\Delta v|^2+|\Delta w|^2}$, is therefore proportional
to $\mu\Lambda_2\sin (\pi/N)$ for small $\mu$.

 For $k=2$ it is clear that the shortest distance between points
separated by two wrappings of $x^{10}$ is given by connecting $A_2$
and $B_2$; the length of the straight line between them is
proportional to $\mu\Lambda_2\sin (2\pi/N)$.  In general the line
which gives $k$ units of flux connects $A_k$ with $B_k$ and has length
$\sim\sin (\pi k/N)$.  The tensions of the $k$-strings thus agree with
the field theory result \tens.

Notice that the $k$-string is pinned near the point where one of the
massless monopoles is localized; any attempt to move the $1$-string
around the ellipse, so that it connected, say, the points $B_2$ and
$B_4$, would give a curve whose length would be of order
$\Lambda_2/R$, which is much greater than $\Lambda\mu$. The positions
of the strings are consistent with our assertion that the $k$-string
is the Nielsen-Olesen string of the $k$-th monopole.  We will now
verify this assertion by looking at quark-antiquark mesons.

\fig{Transition from the Coulomb branch (a) through the massless monopole 
point (b) to \ntwo\ supersymmetry breaking (c). In (a) and
(b) we draw a closed curve $\gamma_k$ in $\Sigma$ which surrounds the
$k$-th D4 brane and wraps once around $x^{10}$. After the transition,
$\gamma_k$ does not exist.  The curve
$\hat\gamma_k$ wraps once around $x^{10}$.  Its segments
$A_k-B_k$ and $B_{k-1}-A_{k-1}$ lie {\it outside}
$\Sigma$, at fixed values of $t$; they can join to a $k$-string
and $(k-1)$-antistring, respectively. }{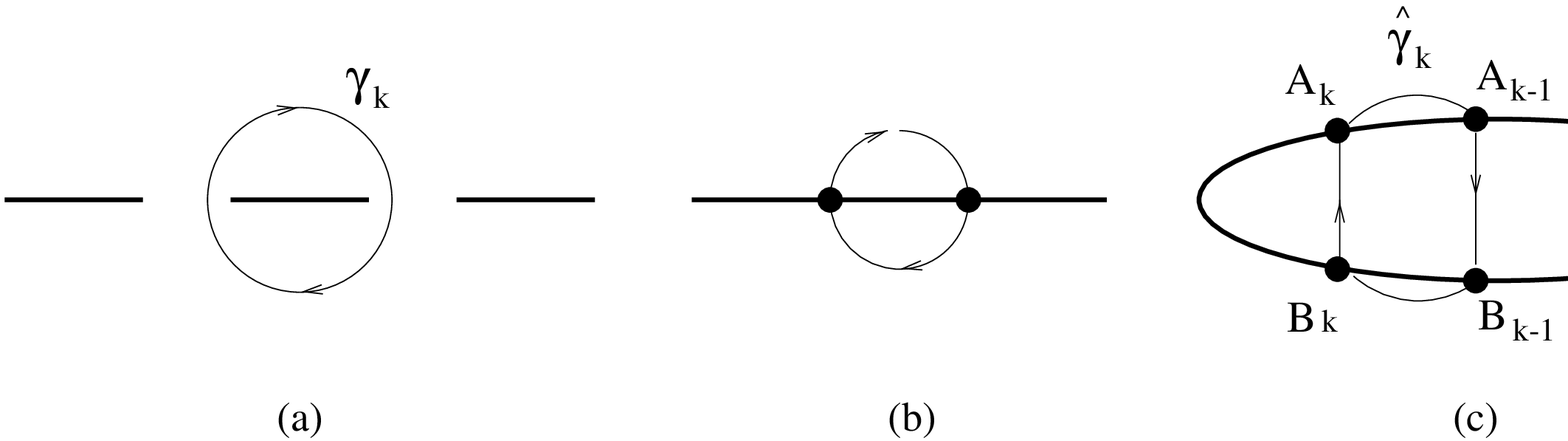}{15 truecm}
\figlabel\coulomb

As discussed previously, far out along the Coulomb branch of the
\ntwo\ theory, a heavy quark $Q^k$ is constructed by attaching one
boundary of a membrane to $\Sigma$ on a curve $\gamma_m$ surrounding
$v=m$, and the other boundary to a curve $\gamma_k$ surrounding the
branch cut at $v=\phi_k$.  As we move in along the Coulomb branch
toward the monopole point, keeping track of the quark $Q^k$, the curve
$\gamma_k$ moves to the position indicated in figure \coulomb.  After
\ntwo\ \susy\ is broken, the curve $\gamma_k$ no longer exists as a
closed curve in $\Sigma$.  It is useful to define a closed curve
$\hat\gamma_k$ which consists of segments
$A_{k-1}$--$A_k$--$B_k$--$B_{k-1}$--$A_{k-1}$, with the property that
the segments $A_{k-1}$--$A_k$ and $B_k$--$B_{k-1}$ lie in $\Sigma$
while the segments $A_k$--$B_k$ and $B_{k-1}$--$A_{k-1}$ do not.  Note
that this curve wraps once around $x^{10}$.  We may therefore
construct a quark $Q^k$ as a membrane with one boundary on $\gamma_m$
and the other on $\hat\gamma_k$, but only if we join it to a
$k$-string along the line $A_k$--$B_k$ and to a $(k-1)$-antistring
along the line $B_{k-1}$--$A_{k-1}$.  These strings can end only by
attaching to an antiquark $\tilde Q_k$.  This reproduces the
results of Douglas and Shenker for mesons, when we identify the
$k$-string as the string of the $k$-th monopole.

It is similarly straightforward to verify that the state $Q^1Q^2\cdots
Q^k$, a membrane which can attach to the curve
$A_0$--$A_k$--$B_k$--$A_0$, connects only to the $k$-string given by
$A_k$--$B_k$.

\fig{The ellipse \curva\ for small $\mu$, with notation as in figure
\ellipse. A $k$-string/$(k-1)$-antistring pair is homotopic to a curve 
stretched along the ellipse and wrapping once around $x^{10}$.  This
curve can be shifted to the position of a $1$-string.  The
intermediate steps involve energies of order $\Lambda_2$
and so this string-antistring partial annihiliation process is
energetically suppressed.}{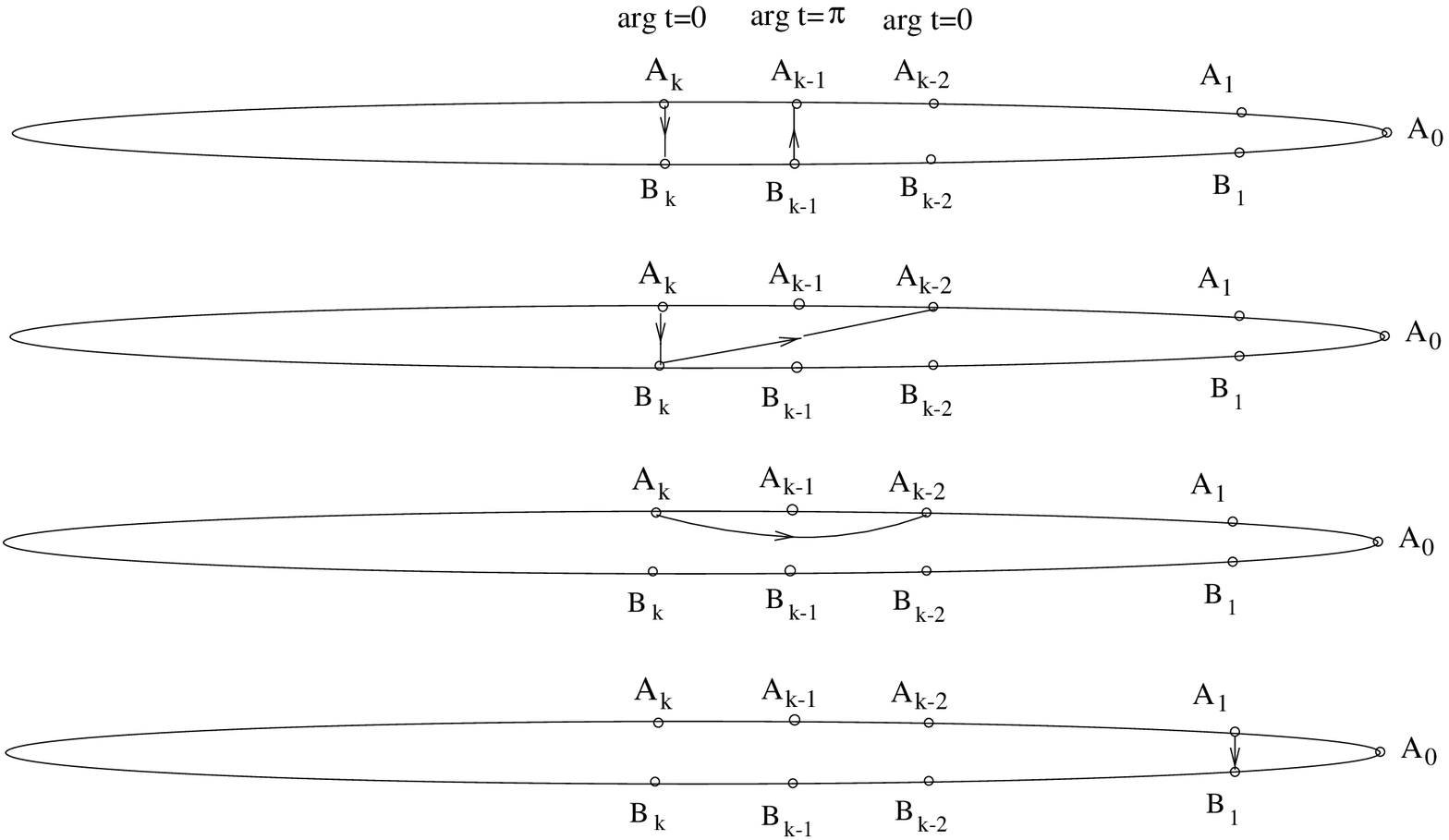}{16 truecm}
\figlabel\deformation

We can also see that sufficiently excited mesons are unstable to $W$
boson pair production.  Ignoring the energetics of the process, this
is easy to see topologically.  Before \ntwo\ is broken, the $W$ bosons
$[W_\alpha]^k_p$ are given by attaching a membrane on the curve
$\gamma_k$ at one end and on $\gamma_p$ on the other.  Here, the $W$
boson can be attached at $\hat\gamma_k$ and $\hat\gamma_p$; on the one
side it will connect to a $k$-string/$(k-1)$-antistring pair, while on
the other it will connect to a $p$-string/$(p-1)$-antistring pair.  In
particular the state $[W_\alpha]^k_1$ can convert the
$k$-string/$(k-1)$-antistring pair to a $1$-string.  Of course, this
process can only occur if enough energy is available, and the required
energy is of order the $W$ mass $m_W\sim\Lambda_2$, which is much
larger than $\Lambda_1$ for small $\mu$.

There is another process, string-antistring partial annihilation,
which is homotopically equivalent to this one.  We illustrate in
figure \deformation\ a deformation by which a
$k$-string/$(k-1)$-antistring pair annihilates to form a $1$-string.
The energy of this process is estimated by multiplying the length of
the segment $A_k$--$A_{k-2}$, of order $\Lambda_2/R$, by the length in
space along the MQCD string which is required for the transition.
Assuming on physical grounds that the required length is of order
$\Lambda_1^{-1}$, and recalling that $\Lambda_1 R\sim 1$, we find that
the energy of this process is again of order $\Lambda_2$.  Without a
detailed study of the energetics one cannot say whether $W$ boson
production or string annihilation is the better language for
explaining the physics; the two descriptions are complementary.

To show the baryons are correctly represented as in figure \dsbaryons\
is completely straightforward, and we omit a detailed discussion.

Now, what happens when $\mu$ is taken large?  Since the $W$ mass is
now less than $\Lambda_1$, the $W$ boson pair production process
occurs rapidly, and the $k$-string/$(k-1)$-antistring pair decays very
quickly to a $1$-string.  A complementary description is that the
string-antistring annihilation process becomes instantaneous.  To see
this, consider figure \deformation\ when $\mu$ is large and the
ellipse is nearly circular.  The line connecting $A_{k-2}$ to $A_k$
will now be shorter than the sum of the lines $A_k$--$B_k$ and
$A_{k-1}$--$B_{k-1}$, so the energy barrier to the annihilation
process is essentially gone.

Thus, for large enough $\mu$, all mesons with quarks in the
fundamental representation are indeed built from a single string
tension, that of the $1$-string.  The $k$-string/$(k-1)$-antistring
pair is no longer metastable.  But the $k$-strings themselves {\it
are} stable.  Their tensions (in M theory) still satisfy
\tensionratio, as a consequence of the geometry of the ellipse \curva\
(figure \ellipse) which goes smoothly over to the circle of figure
\circolo.  These strings therefore are still relevant for
heavy quarks in higher representations, as discussed in section
\multiple. In particular, for quarks in the $k$-index antisymmetric
representation of $SU(N)$, which have charge $k$ under the center of
$SU(N)$, a quark-antiquark meson will be bound by a $k$-string, not by
$k$ $1$-strings.  The same relation suggests that certain excited
baryons may indeed resemble the chain suggested in figure
\dsbaryons.\foot{However, the structure of a baryon will
depend on its excitation quantum numbers; figure \baryons\ may apply
for some states, figure \dsbaryons\ for others, while still others
may have an intermediate structure.}

What are the key physical differences between the large and small
$\mu$ limits?  All of the unusual phenomena can be traced back to the
breaking of the Weyl symmetry and the projection to the $U(1)^{N-1}$
subgroup (abelian projection) implemented by the expectation value for
$\phi$ \DS.

The breaking of the Weyl group is visible in the structure of the
fivebrane.  The Weyl group exchanges color indices, and so in the Type
IIA brane language exchanges D4 color branes.  Since in the vacuum
with massless monopoles the D4 branes are lined up side by side, each
one lying between two adjacent circles in figure \lineseg, it is clear
that the symmetry which would exchange them is spontaneously broken in
this vacuum.  The breaking is still present when $\mu$ is non-zero, as
represented by the inequivalence of the points $A_k$ and
$A_{k'}$. From the positions for small $\mu$ of the curves
$\hat\gamma_k$ (figure \coulomb), it is evident that the quarks with
different color indices are distinguished.  The $\integer{(N+1)/2}$
different mesons result from this breaking.  Only when
$\mu\rightarrow\infty$ does the ellipse become a circle and the Weyl
group become fully restored.

The effect of the abelian projection is that the Nielsen-Olesen
strings are pinned at the points $A_k$ and $B_k$ for small $\mu$,
while as $\mu$ becomes large they break free of these points and move
easily around the ellipse.  This freedom of movement corresponds to
the presence of light non-abelian gauge bosons (relative to
$\Lambda_{1}$) in the theory, in contrast to the small $\mu$ situation
in which the $W$ bosons are all heavy compared to $\Lambda_{1}$.
Having $W$ bosons much more massive than the photons of the unbroken
$U(1)^{N-1}$ gauge theory impedes the free flow of color quantum
numbers, and inhibits annihilation of strings, inevitably resulting in
each $k$-string carrying an approximately conserved quantum number.
Only when the gauge bosons all are treated equally, as in the large
$\mu$ limit, is color free to flow as expected in an unbroken
non-abelian gauge theory. This can be viewed as an argument against
using abelian reduction techniques \tooft\ to study confinement in
QCD.

As a final comment, we note that all of the physics discussed in
this section is essentially trivial in the case of $SU(2)$ gauge
theory.  In this case there is only one meson and only one
type of string for all values of $\mu$, and so the discussion of
Seiberg and Witten \swone\ is the complete story.

\newsec{QCD Strings, Nielsen-Olesen Strings, and BPS Conditions.}

A brief comment on the ``almost BPS'' properties of the Nielsen-Olesen
strings is in order.  As noted in section \review, the massless
monopole vacuum is well-described as an \ntwo\ \susic\ Abelian Higgs
model of $N-1$ decoupled $U(1)$ gauge theories, each with a massless
charged hypermultiplet.  With a properly normalized quartic potential,
such as that generated by a Fayet-Iliopolous term, Abelian Higgs
models have BPS saturated strings \refs{\NielOle,\Bog,\AHvort} in the
semiclassical limit.  The string tension is equal to the coefficient
of the Fayet-Iliopolous term.  Note that
\ntwo\ \susy\ is preserved by a Fayet-Iliopolous term; only the
$SU(2)$ R-symmetry is broken.

As can be seen from \superpot-\uformula, the
breaking parameter $\mu$ acts to leading order in $1/\Lambda_2$ like a
Fayet-Iliopolous term for each $U(1)$ in the monopole Lagrangian.
This results in $N-1$ strings with tension of order $\mu\Lambda_2$.
However, the $1/\Lambda_2$ corrections break \ntwo\ \susy.  A more
serious obstruction, as pointed out in \witMa, is that we
cannot expect \none\ QCD strings to be BPS saturated; it cannot be
that one string satisfies $T\geq Q_{BPS}$ while $N$ identical strings,
which must carry charge $NQ_{BPS}$, can decay to the vacuum.

In exactly what limit does the perturbation of equation \superpot\
become a Fayet-Iliopolous term?  If we take
$\Lambda_2\rightarrow\infty$ and $\mu\rightarrow 0$ simultaneously,
holding $\mu\Lambda_2$ fixed, then the term linear in $a_D^{(k)}$
survives while the non-linear terms drop out.  From the point of view
of the electric description of the theory, this is a rather odd
scaling in which the theory is taken to ultra-strong coupling while
the \ntwo\ breaking parameter is taken to zero.  This is perhaps not
surprising, since on the one hand the Fayet-Iliopolous term must not
break \ntwo\ \susy, and yet it must break $SU(2)_R$.

In short, if $\Lambda_2\gg\sqrt{\mu\Lambda_2}\gg\mu$, then the strings
should behave as though BPS saturated, carrying accidentally conserved
charges and satisfying an approximate Bogomolnyi bound on their
tensions.  The strings' energy scale $\sim \sqrt{\mu\Lambda_2}$ lies
between two difficulties.  Above lie the irrelevant operators which
know the theory is really non-abelian; these violate conservation of
$U(1)^{N-1}$ flux quantum numbers, which would otherwise serve as
conserved BPS charges.  Below we find the relevant but small operators
which break \ntwo\ \susy\ and destroy the associated BPS bound.  All
approximate BPS properties are lost when $\mu\sim\Lambda_2$.  These
aspects of the theory are all clearly visible in the fivebrane
construction discussed in section \softly.

As an aside, we note that the number of Fayet-Iliopolous parameters
always matches the number of \ntwo-breaking terms in \ntwo\ QCD; both
are equal to the dimension of the Coulomb branch, which in turn
equals the rank of the group.  For example, in $SU(N)$, the $N-1$
operators $\tr\phi^2, \tr\phi^3,\cdots,\tr\phi^N$ parameterize the
Coulomb branch.  We may break \ntwo\ \susy\ by adding the general
superpotential
\eqn\generalbreak{
W = \sum_{k=2}^N \lambda_k \tr \phi^k\ ;} 
The $N-1$ Fayet-Iliopolous terms are then linear combinations of the $N-1$
coefficients $\lambda_k\Lambda_2^{k-1}$ in the limit
$\Lambda_2\rightarrow\infty, \lambda_k\rightarrow 0$.  

\newsec{An Additional Comment on Confinement}

It is interesting to consider, in \ntwo\ \susy, what happens to
monopoles when the electric non-abelian gauge group is broken along a
Higgs branch.  We expect the monopoles to be confined by strings
carrying magnetic flux.  In the Abelian Higgs model, this is
well-understood in field theory \refs{\NielOle,\Bog,\AHvort} and a brane
representation (in the context of Type II compactifications) for a
monopole-antimonopole pair bound by an abelian flux tube was given in \gmv. 
In non-abelian \ntwo\ QCD, the Type IIA and M theory brane approach 
that we have been using leads to a related construction.

The simplest theory to consider is \ntwo\ $SU(2)$ with two flavors of
mass $m$.  This theory has a classical Higgs branch at $\phi=m$, which
survives in the quantum theory for $\phi$ sufficiently large.  This
can be seen using branes either by adding D6 branes to the theory or
by adding semi-infinite D4 branes. We choose the latter course, since
we have not discussed D6 branes up to now.

\fig{The Type IIA brane construction for \ntwo\ $SU(2)$ with two doublets.
In each picture, the $SU(2)/U(1)$ monopole is shown as a D2 brane with
boundary on the NS and D4 branes.  On the left, $m>\phi$, while in the
center and right, $m=\phi$.  The low energy theory in the central
picture is $U(1)$ with two massless hypermultiplets.  In the righthand
picture we have moved onto the Higgs branch by moving the infinite D4
brane off the NS branes.  The dark lines indicate edges of the D2
brane which cannot be attached to NS or D4 branes.  A D2 brane extending in
a spatial direction can be sewn onto each of these edges.}
{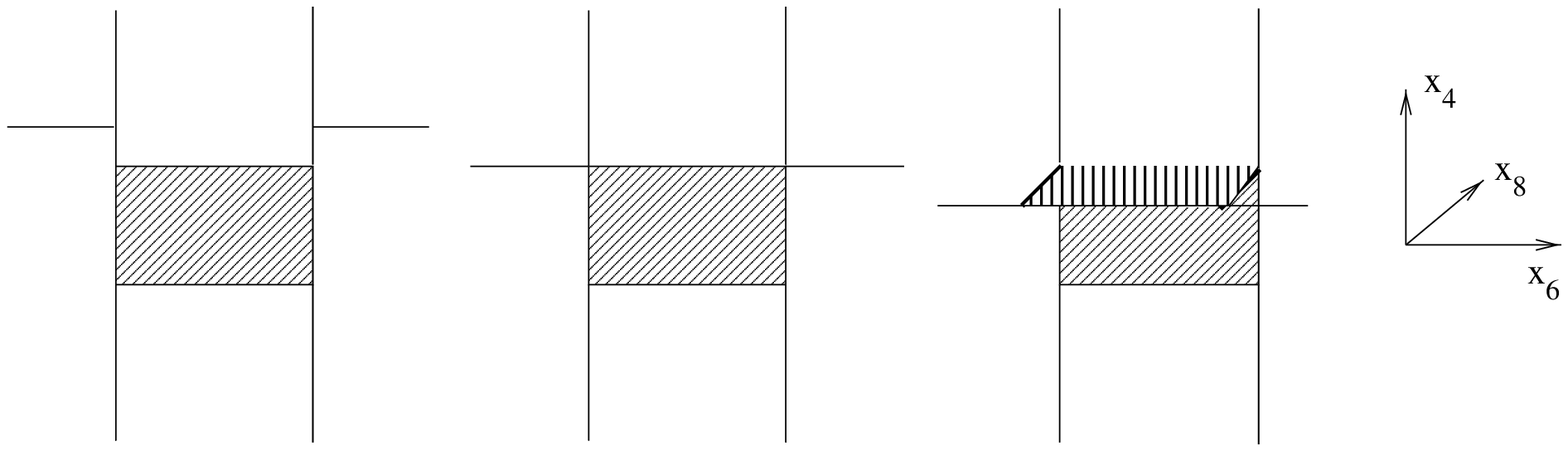}{12truecm}
\figlabel\monoconf

We discuss the question semiclassically, using Type IIA string
language.  Consider $SU(2)$ gauge theory, given by two finite D4
branes between two NS branes, and attach one semi-infinite D4 brane to
each of the NS branes, giving two hypermultiplets in the doublet
representation.  The broken $SU(2)$ theory has a monopole, made of a
D2 brane lying in the shaded region between the color branes, and
attached to the NS branes and the D4 color branes, as shown in figure
\monoconf.  If $\phi=m$ then it is possible to connect the
semi-infinite branes to the finite brane at the same value of $v$ and
pull the resulting infinite D4 brane off of the NS branes, as shown in
figure \monoconf.  This breaks the remaining $U(1)$ symmetry, which
means we have moved along the Higgs branch and expect the monopole to
be confined.  Now, if a D2 brane is attached to the NS and D4 branes,
two pieces of its boundary are left open, as indicated in figure
\monoconf.  This means this configuration by itself is not consistent
--- there are no isolated monopoles.  However, to each of the open
pieces of the D2 brane boundary we may sew on another D2 brane which
also extends along a curve in space.  Because of the relative
orientations of the two boundaries, one of these D2 branes is a string
in spacetime, while the other is an antistring; the pair together are
a magnetic flux tube. The flux tube can terminate by being connected
to an antimonopole (a D2 brane with orientation opposite to a
monopole).   

On dimensional grounds, the flux tube tension ought to be related to
the square of the expectation value by which the gauge group is
broken.  The tension of the flux tube is apparently proportional (in
this semiclassical regime) to the distance between the NS branes and
the infinite D4 brane.  The expectation values $\vev{Q^1\tilde Q_1}=
-\vev{Q^2\tilde Q_2}$ are also proportional to this distance
\hw.\foot{These simple statements about tensions and expectation
values are actually quite naive.  There are both classical and quantum
mechanical subtleties, since this flux tube is not a BPS-saturated
semiclassical soliton; see for example \vortexsubtle.}

The confined monopole-antimonopole state, bound by a string-antistring
pair, is thus a single continous D2 brane.  This structure is very
similar to the picture proposed for the purely abelian case in
\gmv\ and to our earlier quark-antiquark meson.  

We may also consider truly non-abelian weakly-coupled examples.  A
simple example is $SU(3)$ with six equally massive flavors at a point
on the Coulomb branch where it is broken to $SU(2)\times U(1)$ with
six massless doublets, an infrared free theory.  There are monopoles
in the coset $SU(3)/SU(2)\times U(1)$ which are not neutral under
$SU(2)$ (as can be seen by breaking the theory slightly to $U(1)\times
U(1)$.)  On the Higgs branch, where doublet expectation values break
$SU(2)$ completely, the monopoles are confined as described above.

Adapting this mechanism to strongly coupled theories, using the M
theory fivebrane and membrane to replace the NS/D4 and D2 branes, is
straightforward.  Furthermore, a similar mechanism applies for confinement of
quarks along monopole Higgs branches; it is closely related to the M
theory picture of monopole condensation and string formation discussed
in section \softly.

\newsec{Conjectures in Supersymmetric and Non-Supersymmetric QCD}
\seclab\comments

We have seen that in the M theory construction of  \none\ and weakly broken
\ntwo\ QCD, there are stable $k$-strings with tensions satisfying 
\tensionratio.  Witten has proposed a minimal surface fivebrane solution 
to non-supersymmetric MQCD \witMa.  (Other work on non-supersymmetric
QCD using branes has appeared in \nonsusybrane.)  While many aspects
of this solution are not understood, and although not only
renormalization effects but even phase transitions may separate the
semiclassical fivebrane picture from real QCD, we may still try to
construct the MQCD $k$-strings as we have done in the supersymmetric
case.  It is straightforward to verify, as we have done in appendix B,
that for all values of the parameters in Witten's solution, the
candidate $k$-strings still connect points on a curve similar to that
of \curva\ (figure
\ellipse).  Consequently, the formula \tensionratio\ still applies,
and the $k$-strings are stable.

Nevertheless, if we want to extract information about ordinary field
theoretic QCD, we must view a formula such as \tensionratio\ with
considerable skepticism. To what extent can semiclassical results in M
theory give reliable results in strongly coupled gauge theories?

The MQCD gauge theories considered in this paper arise from
configurations of branes in string theory. In the weakly coupled Type
IIA string theory, these configurations realize at low energy the
\none\ and \ntwo\ supersymmetric field theory we are interested
in, but the dynamics of the strongly interacting field theory cannot
be computed there. In the strong coupling limit of Type IIA string theory,
the configuration of branes becomes smooth enough to allow a
semi-classical analysis using M theory on $M^{10}\times S^1$
\refs{\witMb,\witMa}, where $S^1$ is a circle of radius $R\gg 1$ in M
theory units; in this approach much of the non-perturbative structure
of the brane theory may be obtained. However, for large $R$ it is not
obvious which aspects of the field theory physics are in fact
reflected in the MQCD brane theory.

The difference between the \none\ and the \ntwo\ case is simply
understood \refs{\witMa,\witMb}.  The brane configuration depends on
two parameters, the scale $\zeta$ and the radius of the eleventh
dimension $R$. In \ntwo, $\zeta$ fixes the only parameter of the
theory $\Lambda_2$ and we can vary $R$ as we like without changing any
of the physics of interest. In \none, the BPS saturated domain wall
\dvalshif\ and the MQCD string tension depend on $R$ and $\zeta$ in
different ways \witMa.  In order to get a theory which at least
resembles \none\ QCD, we must fix the parameters in such a way that
the MQCD string has a tension of order $\Lambda_1^2$ and the domain
walls of the spontaneously broken chiral symmetry have a tension of
order $\Lambda_1^3$. As shown in \witMa, and reviewed in appendix A,
this requires $R$ to be of order $1/\Lambda_1$. Unfortunately, for
this choice, the Kaluza-Klein modes with momentum around $S^1$, which
of course do not exist in the QCD field theory we want to study, have
masses of order $1/R\sim\Lambda_1$, too low to be ignored.  If we try
to take $R$ small, then the tension of the MQCD string becomes very
large, and the theory does not behave like QCD.
 
It is reasonable to believe that the qualitative properties of \none\
QCD (such as confinement, presence of mesons and baryons, etc.) are
independent of $R$ and are correctly described by MQCD.  By contrast,
quantitative predictions should only be trusted when they are
protected in some way from renormalization, and therefore can be
followed also to small radius.  This is the case for all the
computable properties in the \ntwo\ case and for the domain wall
tension in \none, but is not the case for the QCD string tension.

There are essentially two quantitative results in this paper
concerning \none\ gauge theories. The first is the computation of the
Douglas-Shenker string tensions in the weakly broken
\ntwo\ theory.  We find agreement with quantum field theory,
as is natural, since not only is the theory nearly \ntwo\ \susic,
but also the strings are almost BPS saturated and are described at weak
coupling in the monopole variables. The second result, concerning the
tensions of $k$-strings for strongly broken
\ntwo\ gauge theory, is much more subtle. We can trust our estimate
\klengths\ for the tension of a $k$-string only for large radius 
MQCD. To extend this result to ordinary QCD, we need to be able to
follow this quantity down to small radius. Unfortunately, large
renormalizations are expected, and the tensions implied by \klengths\
certainly cannot be trusted.

However, the fact that all of the fivebrane generalizations of QCD ---
the weakly broken \ntwo, the \none\ and even the non-supersymmetric
case --- exhibit the same {\it ratio} of tensions \tensionratio\ is
remarkable.  Perhaps this is merely a property of semiclassical
M theory, and field theory agrees with it only when it has to, namely
for weak \ntwo\ breaking.  But it is also possible that ratios of string
tensions are rather weakly renormalized and that
\tensionratio\ is fairly accurate for \none\ or even for
non-supersymmetric QCD.  

Even if the formula \tensionratio\ is not quantitatively accurate,
there is still the qualitative question of whether the $k$-strings in
various forms of QCD are stable or unstable to decay to 1-strings,
that is, whether $T(k)/T(1)$ is greater than or less than $k$.  M
theory seems to come down squarely on the side of stability in all of
these theories.  In particular, even though \tensionratio\ certainly
has $1/R$ corrections, as pointed out in the footnote before \points,
these corrections do not alter the stability properties of the
$k$-strings.  We should also note that other techniques, such as the
strong coupling expansion, agree with M theory that $T(k)/T(1)=k$ in
the large $N$ limit; the real issue in this context is the sign of the
leading $1/N$ correction.

The ratios of string tensions can be studied in numerical lattice
simulations of gauge theory.  This would be straightforward for the
non-supersymmetric theory, but could be carried out even for weakly
broken \none\ or \ntwo\ gauge theories \lattsusy.  Of course, for $SU(2)$
or $SU(3)$ there is only one string tension, so one must at least
study $SU(4)$.  To our knowledge this has not been done even in the
non-supersymmetric case \lattrev.  It seems to us that the ratios of string
tensions are fundamental quantities in confining gauge theories, and
that it would be useful to have numerical values for a few non-trivial
examples.

Why should we care whether the M theory result survives, at least
qualitatively, into pure \none\ QCD?  First, it would imply the
presence of markedly different string tensions for mesons built from
quarks in higher representations.  Second, it would confirm our
picture for the transition from \ntwo\ to \none\ QCD, which depended
on \tensionratio\ holding qualitatively. Third, it would give some
indication as to whether M theory is a useful guide for extracting
physics that ordinary field theoretic methods cannot compute.

\newsec{Conclusion}

Extending the results of \refs{\witMa,\Berkeley,\yank}, we have found
that the MQCD picture correctly describes the standard lore for
confinement in \none\ QCD and the field theoretical predictions for
the weakly broken \ntwo\ gauge theory.  In \none\ MQCD we have studied
construction of heavy quark mesons and baryons as M theory membranes,
and identified the $k$-strings of MQCD (a $k$-string is a flux tube
carrying $k$ units of flux, $0<k<N$.)  We have seen explicitly that
the $N-1$ strings of Douglas and Shenker go smoothly to the
$k$-strings of MQCD.  The metastability of a
$k$-string/$(k-1)$-antistring pair explains the presence of the many
quark-antiquark mesons observed by Douglas and Shenker; this
metastability is lost when \ntwo\ \susy\ is strongly broken.  In
addition we have discussed the scaling limit in which the
Douglas-Shenker strings are BPS saturated, and have explored the brane
picture for monopole confinement in \ntwo\ nonabelian gauge theory,
which closely resembles that of \gmv.

We have shown that the \none\ and non-supersymmetric MQCD results of
\witMa\ imply that the ratios of the MQCD $k$-string tensions 
are independent of the \ntwo\ \susy\ breaking parameters.  Our
formula predicts that $k$-strings are always stable.  We have proposed
that this formula might undergo relatively little renormalization and
might hold, at least qualitatively, in ordinary \none\ and perhaps
even non-supersymmetric QCD.  The question of whether these
conjectures are correct could be studied using lattice gauge theory,
and we hope that some attention will be given to this problem.

\vskip 0.3in

\centerline{\bf Acknowledgments}

We would like to thank M.~Alford, P.~Mayr, S.-J.~Rey, N.~Seiberg,
S.~Shenker, F.~Wilczek and E.~Witten for useful discussions.  The research of
A.H. and M.J.S. is supported in part by National Science Foundation
grant NSF PHY-9513835.  The work of M.J.S. is also supported by the
W.M.~Keck Foundation.  The research of A.Z. is supported in part by
DOE grant DE-FG02-90ER40542 and by the Monell Foundation. A.H.
would like to thank the Aspen Center for Physics for hospitality
during the completion of this work.

\appendix{A}{Dimensional Analysis for the M Theory Construction}

We use M theory units in this paper. Quantities are measured in eleven
dimensional Planck length units ($l_P$). 

If we compactify M theory on a circle of radius $R$, we recover the
Type IIA string theory with string length $l_s$ and coupling constant
$g_s$ given by
\eqn\units{R= g_sl_s,\qquad l_P=g^{1/3}l_s.}
The membrane tension is $1/l_P^3$. A membrane wrapped around the
eleventh dimension is identified with the Type IIA string and equation
\units\ correctly gives its tension $2\pi R/l_P^3=2\pi/l_s^2$.

In the M theory picture, quarks and W bosons are identified as
membranes wrapped around $x^{10}$.. The mass of a state corresponding
to a membrane with length $v$ in the $x_4,x_5$ plane is therefore
proportional to $2\pi Rv/l_P^3$.  As a consequence, in the units
appropriate for M theory, the \ntwo\ curve for an $SU(N)$ gauge theory
reads,
\eqn\Mcurve{t + 1/t + \Lambda_2^{-N} \prod_i^N(2\pi Rv/l_P^3 - \phi_a ) = 0.}
where t is dimensionless. In Type IIA units (see equation \units ) the
power of $R$ disappears and the curve assumes the form used in section
\twotheory\ (see for example equation \curvetwo\ with $m=\infty$).

In section \nonetheory, we rotated the \ntwo\ curve and found the
\none\ Riemann surface,
\eqn\Mcurveone{v = {\Lambda_1^{3/2}l_P^3\over (2\pi R)^{1/2}} t^{-1/N}, 
\qquad w = {\Lambda_1^{3/2}l_P^3\over (2\pi R)^{1/2}} t^{1/N}.}
so that $vw=l_P^6\Lambda_1^3/2\pi R$.

\none\ MQCD has  strings \witMa\ with
tension $\sqrt{vw}/l_P^3$ and domain walls with tension
$Rvw/l_P^6$. If we fix the value of $R$ to be $1/\Lambda_1$ we
correctly reproduce the expectations that the string tension is
proportional to $\Lambda_1^2$ and the domain wall tension is
proportional to $\Lambda_1^3$.  We have not determined overall
normalizations.  The exact coefficients could be fixed by a detailed
analysis of the normalization of the superpotential (gluino
condensate) generated by the fivebrane theory.

\appendix{B}{The Non-Supersymmetric Fivebrane Solution}

A non-supersymmetric fivebrane solution reproducing, in a certain
limit, the bosonic $SU(N)$ Yang-Mills theory has been proposed in
\witMa. It is obtained by an arbitrary rotation of the NS$'$
brane.  The equation is parametrized in terms of a complex number
$\lambda$ and depends on two complex four-vectors $\vec p,\vec q$ and
a real constant c, with the constraints
\eqn\hurmo{\vec p\,^2=\vec q\,^2=0,
\,\,\,~~-\vec p\cdot \vec q +{R^2N^2\over 2}(1-c^2)
=0.}  
The condition for unbroken supersymmetry is $\vec p\cdot \vec
q=0$.  Combining $x_4,x_5,x_8,x_9$ in a real four-vector $\vec A$, the
solution looks like (in the notation of \witMa )
\eqn\nonsup{\eqalign{
\vec A(\lambda) &=\RL\left(\vec p \lambda +\vec q\lambda^{-1}\right)\cr
x^6 & = -(RNc)\,\, \RL\ln\lambda \cr x^{10}& = -RN\IM\ln\lambda\cr}}
       
The curve $\Sigma$ still wraps $N$ times around $x^{10}$, and its
topology has not changed, so we can try to find the MQCD string
tension by the same arguments used in section \nonetheory\ and
\physics.  To describe $k$-strings, we need to find the real curve of 
minimum length connecting two points on $\Sigma$ separated by $k$
wrappings around $x^{10}$.  From \nonsup,
the two points must be specified by $\lambda$ values which
differ in phase by $e^{2\pi i k/N}$.  Since $\Sigma$ is
symmetric with respect to exchanging $p$ and $q$, we expect the two
points to lie on the plane of symmetry.  We may take $|p|=|q|$,
without loss of generality, to ensure that the plane of symmetry is
$x^6=\RL\ln\lambda=0$.  The two points must then be given by $\lambda=
e^{i\sigma_1},e^{i\sigma_2}$, with
$\sigma_1-\sigma_2=2\pi k/N$.  Since these two points have the same
$x^6,x^{10}$ coordinates, the length of the line connecting them is
given by
\eqn\diff{\eqalign{
\Big|\vec A(e^{i\sigma_1})
 -\vec A(e^{i\sigma_2})\Big| &= 
\Big|-2\sin{\sigma_1-\sigma_2\over 2}\IM\left 
(\vec pe^{i\alpha} -{\vec q}e^{-i\alpha}\right ) \Big|\cr
&=\sqrt{2}\sin{\pi k \over N}
\left ( |p|^2+{|q|^2}
+2\RL \vec p\cdot\vec q -\vec p\cdot\vec q^*e^{2i\alpha} -\vec
p^*\cdot\vec q e^{-2i\alpha}\right )^{1/2} \ ,  \cr}}
where $\alpha$ is $(\sigma_1+\sigma_2)/2$.  This is minimized at
$\tan\alpha=-\IM\vec p\cdot\vec q^*/\RL\vec p\cdot\vec q^*
$ for all values of $k$.  We thus find the string tensions
are proportional to $\sin\pi k/N$, as claimed in section \comments.

\listrefs
\end